# Framework, Standards, Applications and Best practices of Responsible AI : A Comprehensive Survey

Thippa Reddy Gadekallu[a], Kapal Dev[b], Sunder Ali Khowaja[c], Weizheng Wang[d], Hailin Feng[a], Kai Fang[a], Sharnil Pandya[e], Wei Wang[f,g]

[a]*College of Mathematics and Computer Science, Zhejiang A&F University, Hangzhou 311300, China*
[b]*Department of Computer Science and ADAPT Centre, Munster Technological University, Bishopstown, Cork, T12 P928, Ireland*
[c]*School of Computing, Faculty of Engineering and Computing Dublin City University, Glasnevin Campus, Dublin, Ireland. D09 V209.*
[d]*Department of Computer Science, City University of Hong Kong, Hong Kong SAR, China*
[e]*Department of Computer and Information Sciences, Faculty of Engineering and Environment, Northumbria University, Newcastle upon Tyne NE1 8ST, United Kingdom.*
[f]*Guangdong-Hong Kong-Macao Joint Laboratory for Emotion Intelligence and Pervasive Computing, Artificial Intelligence Research Institute, Shenzhen MSU-BIT University, Shenzhen 518172, China*
[g]*School of Medical Technology, Beijing Institute of Technology, Beijing 100081, China.*

**Abstract**

Responsible Artificial Intelligence (RAI) is a combination of ethics associated with the usage of artificial intelligence aligned with the common and standard frameworks. This survey paper extensively discusses the global and national standards, applications of RAI, current technology and ongoing projects using RAI, and possible challenges in implementing and designing RAI in the industries and projects based on AI. Currently, ethical standards and implementation of RAI are decoupled which caters each industry to follow their own standards to use AI ethically. Many global firms and government organizations are taking necessary initiatives to design a common and standard framework. Social pressure and unethical way of using AI forces the RAI design rather than implementation.

*Keywords:* Responsible AI, AI Ethics, Standard Frameworks, Global Standards, Implementation Challenges, Social Impact.

## 1. Introduction

The actual beginning of Artificial Intelligence (AI) started in the 1950s by Alan Turing who asked the question "Can machines think? "[1]. In 1956, John McCarthy coined the term 'Artificial Intelligence', later researchers created the first ever running AI program. Neural networks with backpropagation algorithms are widely used AI applications [2]. Later in 2010, data science and deep neural networks such as the convolutional neural network had visibility. Later in 2022, development in Large Language Models (LLM) such as OpenAI, ChatGPT, and Generative AI attracted enterprise value and made impactful changes in the performance of AI. AI has become a vital part of computer science due to its productivity gains [3]. Markets and industries rush to invest in AI to benefit from it [4]. However, the other dark side of AI [5] is that ambiguity exists in AI models, biased nature, lack of transparency [6] and abstract nature of standards. Hence, there is a high social pressure to implement trustworthy AI that benefits the society in an ethical way. Many governments and international organizations try to promote AI standards which document the standards and guidelines to use the AI from a legal and ethical point of view [7].

AI tremendously contributes to the accelerated momentum of technology development that paves to lot of opportunities which will bring numerous social and human benefits. AI has made a drastic transition in industries which solves most of the real-world challenges in the market. This tremendous growth in AI technology carries a deep responsibility to build AI that benefits society. The various challenges involved in developing AI-based projects are discussed in various forums by the researchers to focus on solving the societal requirements, mitigating the risks, and analysing the ethical issues. Several steps are taken by many international and national organizations, industries, and academic organizations to instigate the ethical usage of AI.

Responsible AI (RAI) is a framework to manage risks and challenges associated with AI-based applications. It's a high time to evaluate and analyze the existing AI practices to develop a responsible ethical AI that aligns with the updated regulations. In general, RAI is defined as using AI responsibly that helps in the development of human society constructively. Some of the key principles of RAI are listed as follows:

1. Transparency: AI systems should be transparent in their decision-making processes and provide explanations for their actions to users.

*Email addresses:* thippareddy@ieee.org (Thippa Reddy Gadekallu), kapal.dev@ieee.org (Kapal Dev), SunderAli.Khowaja@dcu.ie (Sunder Ali Khowaja), weizheng.wang@ieee.org (Weizheng Wang), hlfeng@zafu.edu.cn (Hailin Feng), Kaifang@ieee.org (Kai Fang), sharnil.pandya@northumbria.ac.uk (Sharnil Pandya), ehomewang@ieee.org (Wei Wang)

2. Accountability: Developers and users of AI systems should be held accountable for the outcomes of their technology and take responsibility for any harm caused.

3. Fairness: AI systems should be designed and implemented to promote fairness and prevent bias or discrimination against individuals or groups.

4. Privacy: AI systems should respect and protect the privacy of individuals by ensuring that personal data is handled securely and used only for its intended purpose.

5. Robustness: AI systems should be designed to be reliable and robust and capable of performing consistently and accurately in various conditions.

6. Safety: AI systems should prioritize the safety of users.

Researchers have done a survey on RAI that focuses on refining the standards and guidelines. Authors have done a quantitative analysis on AI that focuses on guidelines to apply RAI, and the role of external factors to implement RAI [8]. In this paper [9], researchers have discussed the challenges of using RAI practically in the industries like abstract nature of guidelines, quantitative metrics to measure the accuracy of RAI, and solutions for the listed challenges. This paper [10] argues on the notion of labeling AI. It is better to focus on implementing X on AI rather than on XAI as it would induce the common people to misattributions wrong targets. Authors in [11] focus on industry can contribute to these standards. Researchers in [12] have discussed challenges in implementing RAI models over foundational models. Based on the existing survey, the broad dimension of the current status of standards, design, and implementations is broadly explained in our survey paper. The motivation behind this survey is as follows:

- To promote AI research and development towards socially responsible applications.
- To mitigate the risks involved in using AI systems.
- To investigate the legal and ethical issues that persist with AI.

Although RAI ensures the usage of AI systems ethically, there is a huge gap between the defined guidelines for RAI and the implementation of RAI. This survey paper elaborates on the various dimensions of RAI on its benefits, challenges, and solutions. The main contributions of our paper are as follows:

- Discusses the frameworks, standards and regulations, ethics, and principles articulated for designing and implementing RAI by different industries and organizations.
- Summarizes the current ongoing projects, applications of RAI in various domains and integration with different technologies.
- Analyses the challenges of the socio-technical nature of RAI and the solutions and best practices to overcome its decoupled nature.

The rest of the survey paper is organized as discussed below. Section 2 presents different AI frameworks and highlights the milestones of trustworthy AI progression. Section 3 and Section 4 deal with RAI principles, design, and implementation of RAI practices followed in industries and projects. Organization level, product level and global level AI standards and Regulations are listed in Section 5. Applications of RAI in different industries are discussed in Section 6. Critical analysis of using RAI with various technologies is discussed in Section 7. Current ongoing research and industry projects are discussed in Section 8. Some of the challenges and best practices to overcome these challenges are presented in Section 9. Finally, Section 10 concludes the survey paper by providing an overview and suggestions for using ethical and safe AI.

## 2. Trustworthy AI Framework

With the recent advancements in the field of artificial intelligence (AI), a unified opinion from academics, industries, and policymakers has emerged, which is to develop "trustworthy AI". The stress of incorporating trustworthy AI has been looming for a couple of years now, as evident by numerous regulatory proposals and policy frameworks [13, 14]. Several private companies have also emerged in the form of auditing firms that lend support to their clients for making their AI systems trustworthy [15]. The main aim of trustworthy AI is to induce trust in people for the AI technologies so that more users rely on AI services, hence unlocking the social and economic potential. However, trust cannot be developed overnight, it takes time to design principles, protocols, standards, and practices for trustworthy AI. Therefore, developing a trustworthy AI framework is the most promising aspect for technologists and regulators to seek trust from a broader community for the uptaking of AI [16]. Some researchers have indicated that the design of trustworthy AI frameworks from individual researchers, institutions, or technologists would not be enough to build trust in general, as the framework can be biased towards the community that is historically discriminated [14]. In this regard, a framework that is developed by respectful organizations, governments, and standardization bodies supporting people from diverse backgrounds can be a starting point for building trust [17]. This section provides details regarding the background towards trustworthy AI inception and progression, and existing frameworks towards trustworthy AI.

### 2.1. History of Trustworthy AI

The inception of trustworthy AI came into being by the European Commission following the announcement of their AI strategy [18, 19] in 2018. The idea was to focus on a regulatory and ethical AI framework [20]. In this regard, an expert group was tasked with developing ethical guidelines for AI projects. This group was named as High-Level Expert Group (HLEG) on AI. The group comprised of over 50 experts on subjects from diverse field of expertise and sectors [21]. The group underwent an iterative and comprehensive process to develop AI governance principles for the EU. Most of the process for developing the principles have not been disclosed to the general public except for the HLEG meetings. However, the institutional and the public were able to interact with HLEG through AI Alliance [22, 23] platform, which was used to get feedback on the ethical guidelines proposed by the group. The first draft of their ethical guidelines focuses on the list of use cases and assessments that can be employed by users, deployers, and developers for operationalizing trustworthy AI, use of non-technical and technical methods to realize trustworthy AI, and the development of AI systems should be based on principles and values compliant with human rights laws and other associated charters [24]. The revisions were made on ethics guidelines for trustworthy AI based on public feedback and were made public in 2019 [25]. This was the first ever document with agreement and conceptual understanding of using AI within the European Union. The document was in accordance to the Charter of Fundamental Rights of the European Union and was customized to the European Union fundamentals and values, it was nevertheless a stepping stone towards a comprehensive trustworthy AI framework [26]. The ethical guidelines from HLEG can be summarized in three parts, which are outlined below.

- The design of AI should adhere to robustness from a social and technical perspective, such that cybersecurity, attack resiliency, reliability, accuracy, and functionality should be considered while catering to social responsibilities towards the environment and societal processes [27].
- The design of AI should adhere to ethical guidelines for trustworthy AI proposed by the group.
- The design of AI should adhere to legal obligations.

Considering the three cases outlined above, the conceptualization of such a framework is a daunting task in itself. For instance, adhering to ethical guidelines alone should meet the standards in the areas of accountability, environmental and societal well-being, fairness and non-discrimination, diversity, transparency, data and privacy governance, safety and technical robustness, and human agency and oversight [28]. The said standards themselves were derived from the core principles of explicability, fairness, prevention of harm, and respect for human autonomy [29]. The HLEG's document ethical guidelines for trustworthy AI incorporated the core principles and standards derived from the feedback provided by AI Alliance.

Building upon the HLEG ethical guidelines, the spark to develop a more comprehensive trustworthy AI framework was ignited by the release of public generative artificial intelligence models in late 2022. As the general public was fascinated by the advancements in AI, some researchers and policymakers highlighted the problems associated with generative AI-based methods [13, 30, 31] such as trustworthiness, sustainability, security, digital divide, and safety. We provide a brief timeline in Figure 1 to highlight the milestones for trustworthy AI progression. In continuation to HLEG and based on the progression of Autonomous AI, the first European Union Artificial Intelligence Act (EU AI Act) was proposed in April 2021 [32]. In January 2023, the National Institute of Standards and Technology (NIST) developed the Artificial Intelligence Risk Management Framework on the directives of Congress to help organizations adopt trustworthiness into their evaluation, use, development, and design of AI systems, services, and products. In October 2023, the Hiroshima Process International Code of Conduct for Organizations Developing Advanced AI systems was released which provided voluntary guidance to organizations developing AI systems to promote trustworthy, secure, and safe AI [24]. In October 2023, an executive order was issued by the White House to federal agencies and congress for developing policy and technical tools for trustworthy, secure, and safe development of AI by leveraging privacy enhancing technologies (PET) [28]. The order was issued to combat and protect privacy risks that might arise from the broader usage of AI. In November 2023, the guidelines for secure AI system development by NCSC were issued. The guidelines highlighted the issue of security in the development and overall lifecycle of AI. The guidelines also recommended using PETs for mitigating risks associated with AI systems. In November 2023, global leaders from 28 countries gathered at the AI Safety Summit and signed the Bletchley Declaration to signify the importance of responsible and trustworthy AI, which could help to preserve and protect data privacy [21]. In February 2024, the U.S. government gathered leaders from academia, civil society, and industry and created the NIST AI Safety Institute Consortium to support the deployment and development of trustworthy AI in order to protect the innovation ecosystem. In March 2024, the EU AI Act was approved by the European Union, which provides the world's first set of regulatory principles to govern the use of AI, and ensures the protection and privacy of personal data in the AI ecosystem through their guidelines [32]. In March 2024, a U.S.-led AI resolution was adopted by the United Nations General Assembly [33], which is a standalone resolution to encourage AI governance in the member states to promote trustworthy, secure, and safe adoption of AI systems. In May 2024, AI principles were updated by the Global Economic Policy Forum OECD [34] for developing a trustworthy AI framework that would regulate the lifecycle of AI for maintaining reliability, transparency, and security concerning the management and usage of personal data. In May 2024, the United States Senate proposed the roadmap for AI policy [35] that prioritized transparency, security, interoperability, explainability, and trustworthiness when harnessing the power of AI for system designs and development. As of writing this work, many such developments are being proposed for the design of trustworthy AI systems, guidelines, and principles, accordingly [31, 20].

*2.2. Trustworthy AI Frameworks*

Generally, there is not a unified agreed-upon definition and characteristics of Trustworthy AI, however, different organizations have laid out frameworks to characterize the attributes of Trustworthy AI within the context of public policy, markets, economy, education, and engineering [30, 20, 19]. The organizations that have proposed the characteristics of Trustworthy AI include but are not limited to the National Institute of Standards to manage and evaluate the risks associated with the AI system deployment in the context of the digital environment. The AI Risk Management Framework (ARMF) was announced by NIST in response to preserve the civil freedom, rights, and privacy of users who want to use AI systems. The motive of NIST was to reap the benefits of AI technology while addressing its drawbacks. The ARMF by NIST provides a systematic way to assess, evaluate, and recognize the hazards associated with AI systems. The ARMF considers the following variables to determine the system's reliability, i.e. Validity and Dependability, Security and Resiliency, Improved Privacy, Transparency and Accountability, Interpretibility and Explicability, and Fairness with zero negative biases.

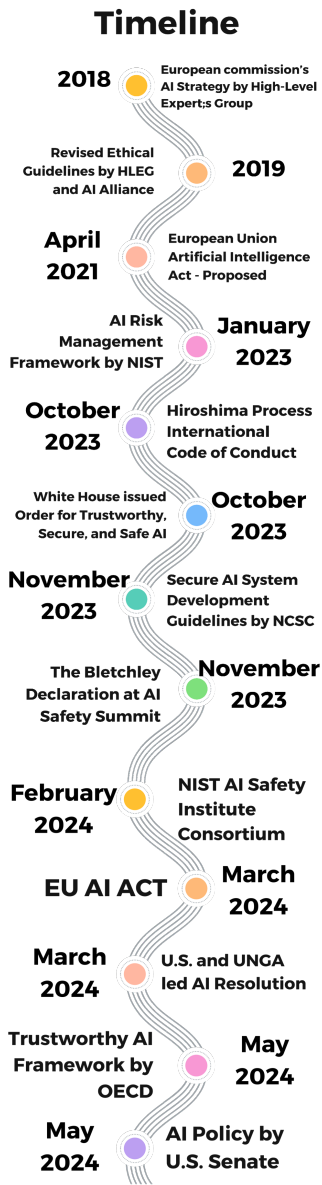

Figure 1: Trustworthy AI History and Timeline

The validity and dependability of an AI system refers to the ability of functioning as per the desired requirement and consistently generating the output with acceptable accuracy. The security and resiliency refers to the ability of AI system for exhibiting strong security against unauthorized access, misuse, and malicious attacks. Furthermore, it also considers the ability to recover from disruptive incidents. The improved privacy aspect refers to the ability of AI system preserving user's privacy and safeguarding sensitive data while adhering to the data protection rules and laws.

The transparency suggests that the system's working should not be concealed as black-box rather it should be be understandable to humans. On the other hand, the accountability refers to the system's auditability and suggesting that there should be someone who bears the responsibility of acts committed by AI systems. The interpretability and explicability refers to the systems ability of providing rationale and justifications behind the actions or decisions. Lastly, the fairness with zero negative biases suggests that the AI system should not incorporate any kind of discrimination or unfair biases.

The ARMF by NIST assesses the risks associated with the AI system and provides practical advice so that the risks can be reduced within the the AI system's lifecycle. The practical advice through ARMF are generally divided into four phases, i.e. govern, mapping, measure, and manage. The govern phase provides suggestions concerning the organization's mission, values, culture, and risk tolerance. The suggestions ensure that the assessments and compliance with respect to the ARMF should be integrated in organization's mission, vision, and policies. Mapping gathers information from all the stakeholders with respect to the potential risks involved in the deployment of AI systems. The gathered information is then translated to the mapping process for identifying, evaluating, and addressing the possible sources or risks. The measure function provides recommendations for methodologies, tools, and techniques that can be used for monitor the risk, as well as analyze and assess the level of risk associated with AI system. Furthermore, the stage also recommends documenting the system functionality, trustworthiness, and social impact. Lastly, the manage function suggests to allocate resources in accordance with the identified risks, and information gathered from previous phases.

and Technology (NIST) [36], United Nations Educational, Scientific and Cultural Organization (UNESCO) [37], The Institute of Electrical and Electronics Engineers (IEEE) [38], and Organization for Economic Co-operation and Development (OECD) [34]. Each standard-setting organization defines the Trustworthy AI traits in a different context. For instance, UNESCO limits the characteristics defining focus to human rights and education, while the IEEE defines the Trustworthy AI traits from robust standards and engineering applications perspective. In this section, we briefly define each of the frameworks and then provide a comparison between the frameworks in terms of their contextual characteristics.

*2.2.1. NIST*

One of the popular Trustworthy AI frameworks was proposed by NIST in January 2023 [39]. It is an AI Risk Management Framework that comprises industry-neutral guidelines

*2.2.2. UNESCO*

Since the inception of chatGPT (a generative AI approach), that can be used by general public, the initial concern was raised by the education system whether the use of generative AI is appropriate or not as it can be used for complete the assignments and solve problems without understanding the underlying concepts, hence undermining the value of understanding, learning, and assessment. In this regard, UNESCO came up with the "Recommendations on the Ethics of the Artificial Intelligence" [40] which helps to promote the diversity, culture, gender equality, inclusion, human agency, and freedom of expression. Although the UNESCO's framework was provided in 2021, it got more attention post the announcement of chatGPT and similar generative AI tools. Some studies have extensively highlighted the problems concerning the use of generative AI such as chatGPT and education system [13].

The framework proposed by UNESCO addresses issues concerning generative AI in the context of research and education. The idea of this recommendation is to prioritize human-centered approach, while using the technologies to support the human development and capabilities. The initial version of the UNESCO recommendations were centered around protecting cultural diversity and human dignity by introducing regulations for AI systems. UNESCO suggested that the AI systems should ensure public accountability, transparency, and human agency. UNESCO refined its policies in 2022 [41] by extending the same human-centered approach towards the risk and benefits of AI in education sector. Concrete recommendations were designed for:

- developing skills and understanding for meaningful and ethical use of AI.
- Monitoring learning processes and alerting teachers in the case of risks and failures.
- Improving quality and expanding access to learning while improving upon data-based provisions.
- Support open-learning and personalized options.
- Enabling the learning programs to be inclusive and accessible to vulnerable groups, such as the ones with disability.

Post chatGPT, the UNESCO policies were reviewed and mapped to the creative use of generative AI in April 2023 [42]. A total of seven steps were laid out to reassert public order and regulate the use of generative AI across different sectors, including education. The seven steps are defined below.

- Step 1: Develop national or endorse international general data protection regulations.
- Step 2: Revise/Adopt and fund whole-of-government strategies on AI
- Step 3: Implement and Solidify regulations for AI ethics
- Step 4: Enforce, Adjust, and Regulate Copyright laws for the content generated using AI
- Step 5: Regulatory frameworks on generative AI should be elaborated.
- Step 6: Capacity building for proper use of generative AI in research and education
- Step 7: Reflect on long-term implications of generative AI for research and education

The report/recommendations by UNESCO also provide separate guidelines to regulate separate guidelines for generative AI for government regulatory agencies, providers of generative AI tools, institutional users, and individual users, accordingly. Currently, in accordance with the generative AI evolution, the working groups associated with Human and Social Sciences Sector Services of UNESCO that include businesses, governments, academic establishments, experts, civil society, and NGOs, are reflecting upon the facts from the 2023 version of UNESCO's recommendations [42]. The working group is currently conducting discussions on several forums including "the Women 4 Ethical AI network", "AI Ethics Experts without Borders network", and "The first global forum on Ethics of Artificial Intelligence" to revise the framework, respectively.

*2.2.3. IEEE*

The institute of electrical and electronics engineering (IEEE) also designed a framework for the ethical use of AI first in 2016 [43, 21], and then revised it in 2019 [44, 20]. The IEEE framework mostly focuses on the research aspect of the AI ethics rather than the industrial aspect. The first AI ethics framework by IEEE was named as Ethically Aligned Design principles, which was prepared by integrating the evaluations from more than 100 global AI and ethics experts. In their second iteration, ethically aligned design (EAD) focused on the "Vision for prioritizing human well-being with autonomous and intelligent systems", which was a global initiative to lay down a set of conditions that could build trust on autonomous and intelligent systems, which include Effectiveness, Competence, Accountability, and Transparency [24].

The effectiveness aspect assumes that the systems do not work exactly as they are promised. However, it should generally fit the purpose of the system, which is the main motto of effectiveness. The evaluation of system effectiveness includes testing and validation protocols during development as well as deployment stages [26]. Compliance with certifications and technical standards. The risk assessment of the algorithm being used as it should not exhibit harm when the system is deployed in real-world situations [27]. The impact evaluation of the algorithm which assesses the impact of the deployment on the subjects, and benchmarking studies. The effectiveness aspect also suggests that the performance measures should be comprehensible to all the stakeholders in order to build trust with the system. Thus, it is recommended that the system should undergo evaluation, not only by the experts who designed it, but also by forensic analysts, who can evaluate the effectiveness objectively [28].

The competence aspect is also concerned with the sound evidence of the AI system but for those who are involved in deployment and development of AI systems. As the report states "A credible adoption of AI into law requires the development of professional measures of competence in the use of AI" [29]. Thus, it is important that the ones who are evaluating or assessing the deployed AI system should have the necessary skills such as conformity of assessments, demonstrations of accuracy, years of experience, certifications, and so forth. One of the examples of such competence aspect can be identified when large discrepancies from the participants were observed in their actual and estimated recall values at Legal Track of the NIST 2011 Text Retrieval (TREC).

The accountability aspect suggests that the developers and designers of the system should be held accountable if their designed system exhibits harmful output characteristics [32]. The accountability aspect suggests including a comprehensive map of responsibilities and roles in the AI design lifecycle so that the responsible individual could be traced back. Additionally, it suggests that sound documentation is established that provides explanations to ordinary citizens, judges, lawyers, outside auditors, and internal compliance departments regarding their systemaˆ€™s inner workings and deployment [31].

Lastly, the transparency aspect suggests that the developers should make the system transparent, which can be characterized into two elements:

- Access to the information regarding the deployed AI system, such as its training and validation procedures, machine learning algorithms, and testing methods used for the design [34].
- The information that can be accessed by different audiences to make them understand why the system behaves differently under hypothetical circumstances [35].

The transparency aspect provides explanatory information on why the AI system takes certain decisions while providing explanations of the system modules to all types of stakeholders. Furthermore, the transparency aspect also caters to both non-experts and experts such that the non-experts should not be offered pseudo-scientific jargon rather they should be provided with sound evidence [33]. The EAD framework puts a lot of emphasis on the transparency aspect and suggests the developers of autonomous and intelligent systems adhere to the conditions of transparency and other trust conditions.

*2.2.4. OECD*

The organization for economic co-operation and development (OECD) was mainly established to finance Europe's reconstruction after the second world war. Since then, the OECD has focused on economic collaboration among the 38 member countries. OECD also proposed an influential set of principles for the ethical use of AI and policies concerning trustworthy AI in 2019. The principles proposed by OECD were named as the OECD framework for classifying AI systems. Unlike UNESCO and IEEE, OECD's aim for defining AI principles is mainly targeted at the economic sector [20]. The goal of the OECD framework is to assess risks, assist with risk management, support sector-specific frameworks, inform data inventories, and understand AI systems to eliminate potential risks in the form of robustness, explainability, and biases [21]. The principles in the OECD framework include:

- Accountability: This principle focuses on privacy and accountability aspects such that the designed AI system should function properly and responsibly throughout its entire lifecycle while adhering to the privacy of all stakeholders [34].
- Safety, Security, and Robustness: This principle focuses on consumer security and ensures that AI systems are resilient to digital attacks and do not exhibit unreasonable risks [24]. To ensure security, the AI system should maintain a record of processes involved in data cleaning, data sources, and data characteristics, which also maintains traceability [32]. The AI system should also consider risk management strategies and maintain documentation concerning risk-based decisions.
- Explainability and Transparency: This principle focuses on the transparency of AI systems such that information regarding training processes, operability, building blocks, and input data is available [27]. This information will help consumers make informed choices about the system. Explainability in this principle helps consumers understand what led the AI system to reach a particular conclusion [26].
- Fairness and Human-centered values: This principle is designed to gain public trust in AI systems by placing privacy, social justice, rule of law, fairness, equality, and human rights at the center of AI system development and functioning [28]. Failure to adhere to this principle may lead to human rights infringements and discrimination [35].
- Well-being, Sustainable Development, and Inclusive Growth: This principle recognizes the potential of AI technology and thus expects AI systems to contribute to sustainability, society, and humanity [33]. The principle also suggests that AI systems should mitigate risks/negative impacts and social biases in decision-making while empowering everyone.

Some tools are available that help adhere to the OECD framework in the form of a checklist. The checklist ensures that the right set of policies is used to validate each of the aforementioned principles and that gaps are addressed in the context of the deployed AI system and the deploying organization [34].

*2.2.5. Summary*

The summarization of the Trustworthy AI frameworks discussed above is provided in the Venn diagram shown in Figure 2. The NIST framework is mostly focused towards risk management, where as the OECD framework is a market-oriented framework and UNESCO is a designed around human-rights framework. Many of the aspects in these frameworks are common, such as safety, security, interpretability, resiliency, accountability, transparency, explainability, and fairness. Similarly, the OECD and UNESCO have common recurring aspects of sustainability and human priority, while the NIST and UNESCO have common aspect of Privacy. There are some unique aspects to each of the frameworks, however, what differs the frameworks are the market forces they cater to. Therefore, even common aspects will be applied and evaluated differently when put into practice. It should be noted that we have not used the IEEE framework in the Venn diagram comparison, the reason for not including IEEE framework is the lack of unique aspects in comparison to the other frameworks.

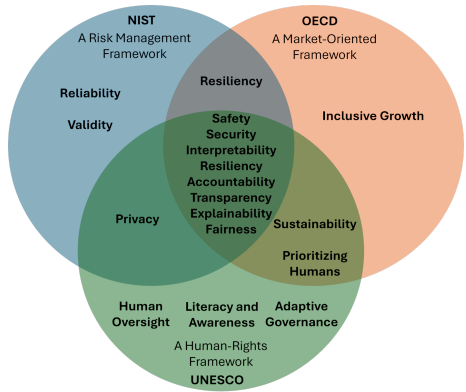

Figure 2: A Venn diagram of trustworthy AI frameworks highlighting the unique and common aspects.

## 3. Designing and Implementing RAI practices

RAI deals with using AI in a legal and ethical manner to enforce AI in a safe, trustworthy, and standardized fashion [20]. Deploying RAI would increase transparency and reduce AI bias [21]. Practicing RAI ensures fairness, transparency, and reliability [27]. AI standards and policies need to be common, and they should depend on data scientists and software engineers who develop AI models in their company [34]. Therefore, the ethical framework for the design and implementation of RAI varies from company to company.

RAI is still an emerging area of AI governance where the word responsible focuses on both ethics and AI democratization, which means the development of AI applications by providing pre-built algorithms, user-friendly interfaces, and cloud computing platforms [33]. This would help non-experts access and develop their own AI and machine learning models [35].

In general, the design and implementation of AI systems differ per organization as principles and policies vary accordingly. The most challenging task in the industry is to provide an unbiased, standardized, and trustworthy AI model for designing and implementing AI systems in organizations [45]. Resources required to build the AI system should focus on the following parameters: shared code repositories, approved variables, standard testing tools for AI systems, stability standards to ensure the working principle of AI programming as designed, and approved model architectures to build AI models. Any organization should implement RAI using the following guidelines:

- Data should be simple and easy to interpret by humans.

- Documentation prepared for any AI-based project should be in such a way that it can be reverse-engineered if there is any mistake in the design process [28].

- Constructive discussions and diverse work culture need to be promoted to avoid bias in project development [26].

- Human-understandable data needs to be presented with the help of interactive features [32].

- A rigorous development process needs to be created to provide visibility to the applicationˆ€™s latent features [24].

- The black-box AI development model needs to be avoided. AI engineers can focus on building a white-box AI model to provide an explainable AI system for each decision made by an AI system.

## 4. RAI Ethics and principles from Industry perspective

Generally the data sets used to train ML and AI models will face the bias issue which occurs due to incomplete and errors in the data. This results in biased AI model which will affect the credibility of the AI model. For example, cases like rejecting applications for financial loans or inaccurate patient diagnosis are due to data bias in AI models. Similarly for developing software programs with AI features, there may be bias issue which can produce an inaccurate result. So, to overcome the above said issues, a standard, transparent and trustable RAI need to be framed to reduce the AI bias.

### 4.1. Microsoft

Microsoft has designed its own RAI governance framework with the help of two AI groups such as Ethics and Effects in Engineering and Research Committee and Office of RAI (ORA) [46]. Especially, ORA is responsible for setting up AI ethics and standards across all the companies. Through implementation, Microsoft has introduced various RAI checklists, AI guidelines, and templates for various processes in AI systems like Human to AI interaction, Conversational AI, design, implementation, AI fairness checklists, standard data sheets, and AI security [34].

### 4.2. FICO

Fair Isaac Corporation (FICO), one of the leading credit scoring organizations, has framed RAI governance policies for the benefit of employees and end-users [47]. Generally, this organization deals with the programming limitations, fairness, and bias detection of ML models. Data scientists at FICO consider the entire lifecycle of ML models and test their effectiveness and fairness continuously in projects. Finally, FICO has framed the following methodologies for bias detection:

- Explainable models for AI need to be monitored throughout the entire process [27].

- Documenting the working principle of AI models can be done using blockchain techniques [33].

- Creating awareness and sharing explainable AI tools with all employees and customers.

- Rigorous testing to track and detect bias in AI models [32].

*4.3. IBM*

IBM has its own committee to deal with the issues regarding AI systems. IBM AI Ethical Board is a central body that helps in creating ethical and RAI policies within the organization [48]. Some of the guidelines that IBM focuses on are transparency and trust in AI, ethical usage of AI, and open-source resources for the community. Also, IBM has publicly called for AI regulations, but no standards or Acts have been passed [21]. ChatGPT, which works based on generative AI models, has not adopted any specific AI Act or policy. However, the Biden Administration and NIST have published general guidelines for the use of AI. In addition to the Risk Management Framework of NIST, the Biden administration has published the prototype for the AI Bill of Rights and created a roadmap for the National AI Research Resource [35].

*4.4. Accenture*

Accenture company has stated, "Make RAI pervasive and systematic in the Enterprise" [49]. The steps that include achieving RAI in the industry are as follows:

- Establish AI guidelines and principles: Guidelines adopted for RAI should include clear governance and accountability during the design, implementation, and deployment phases.

- Assessment of AI risks: Risks involved in AI systems, applications, and use cases need to be assessed qualitatively as well as quantitatively.

- Testing of RAI systematically: Testing of AI projects is a continuous process involving human impact, explainability, fairness, and safety. Appropriate AI tools can mitigate the problems [24].

- Conduct consistent monitoring and compliance: AI systems must be continuously monitored while executing mitigation and compliance actions [34].

- Sustainability and security management: Compliance actions must be addressed with sustainability, law, privacy, and security.

When industries align with all the above guidelines, they can improve trust and value within their organization and also enhance the retention of experts. AI systems that follow regular compliance practices would build a brand and instill confidence in investors, attracting new customers and increasing profits. Demonstrating ethical practices within a company would motivate talented human resources with a sense of ethical values [33].

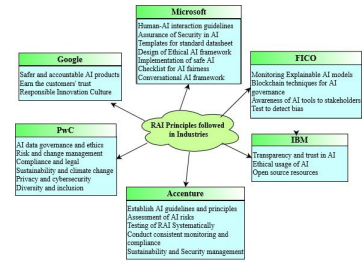

Figure 3: RAI Principles followed in Industries

*4.5. Google*

Google has a Centre of Excellence for Responsible Innovation, a team that guides how to implement AI principles throughout the company [50]. They have planned to use Google Cloud to build advanced technologies and draft policies for their organization. To put these principles into rigorous practice, evaluations are done critically based on two components: deep ethical analyses and opportunity assessments for any technology built within the industry. RAI tools can be used to inspect and understand AI models effectively. Google is building resources like Model Cards, Explainable AI, and TensorFlow open-source toolkits to provide transparent and structured models.

- Safer and accountable AI products: Advanced technologies will be successful only when everyone understands the benefits of using that technology. Building accountable products is crucial to evaluating AI systems.

- Earn the customers' trust: Practicing the best practices in organizations to develop RAI would earn customers' trust to adopt more AI technology [45].

- A culture of responsible innovation: Training AI decision-makers and developers to consider ethical policies while developing AI systems will enable them to find new and innovative missions to the next level [21].

*4.6. PwC*

PwC (PricewaterhouseCoopers) International Limited is a British multinational professional services brand of firms. It is the second-largest professional services network in the world among other accounting firms [51]. The main goal of PwC is to reduce risks, specifically related to risk management and mitigation in AI systems. Risks can arise from different sources, especially wherever AI is implemented. A standard and optimal toolkit is essential to assess potential risks, support mitigation plans, and create an effective AI governance framework [33]. The technical team builds AI systems responsibly and provides innovative ideas across all AI development stages. Focus parameters considered while developing AI projects include AI data governance and ethics, risk and change management, compliance and legal aspects, sustainability and climate change, privacy and cybersecurity, and diversity and inclusion. The sequence of steps involved in building RAI includes assess, build, validate and scale, and evaluate and monitor [34].

## 5. AI Standards and Regulations

AI is a significant technological advancement in today's environment [31]. Based on statistics, international corporations have spent over 306 million on creating AI applications during the last three years. Nine out of ten organizations support AI because it helps businesses become more competitive [46]. However, the challenges in creating and using AI technologies are growing along with the technology. The major challenge is to ensure that AI systems are ethical, dependable, and trustworthy [33]. So, organizations began creating AI standards to make sure the company's ethics, safety, rules, and guiding principles are followed in order to create reliable applications [48].

The creation and promotion of AI standards are currently the focus of numerous organizations. Key organizations involved in development are the International Organization for Standardization (ISO) [34], the Institute of Electrical and Electronics Engineers (IEEE) [44], the European Committee for Standardization (CEN), and the European Committee for Electrotechnical Standardization (CENELEC) [52]. Organizational and product level standards are the two main categories into which AI standards fall. In terms of the governance and regulation of AI, each group fulfills unique but related functions [49].

### 5.1. Organizational Level AI Standards

Organization level AI standards emphasize the broad procedures, rules, and regulations businesses follow to regulate the development, usage, and utilization of AI technologies [34]. In order to ensure compliance with RAI considerations, organizational level AI standards are usually broad in scope and intended to direct the overall AI governance structure inside an organization [33].

Consider the scenario of hiring an employee for the organization. According to organization policy, they seek employees with good qualifications and previous experience [46]. In these kinds of cases, the AI system would look for helpful correlations and use them to make informed conclusions about the hiring process. Following the training phase, a series of examples verify that the system is sufficiently precise and prepared for use. The AI system could, however, become biased if the training dataset is not sufficiently inclusive, balanced, or representative of the problem's dimensions [27]. The algorithm will identify the relationship between gender and acceptability as a kind of bias, for instance, if all employees under scrutiny are connected to men and all rejected applicants are related to women. It will then utilize this discrimination to evaluate new applications. This is unacceptable, particularly when the choices could have an immense impact on people's lives [26].

To address these growing issues, organizations should build a transparent and accurate AI. IBM has released an article towards fairer AI [48]. Establish an efficient AI ethics board first. Next, make the company's AI policies transparent. Collaborate with dependable allies to propel AI ethics forward. Additionally, provide open-source toolkits that enable developers to exchange and obtain cutting-edge datasets and scripts linked to the identification and mitigation of AI bias. This makes it possible for the development community to talk about different ideas of bias and work together.

In addition, make sure your workforce is diverse and that you're funding programs that raise awareness and educate managers, developers, and designers [49]. To determine the best meaning of fairness for your AI, make sure to involve discussions with pertinent social groups and the affected communities. Additionally, develop tools for explainability and transparency to identify bias and its effects [32]. It is recommended that developers establish dedicated, interdisciplinary teams, provide their staff with proper training, and clearly delineate who within the company has responsibility for the outcomes of their AI systems. To overcome AI challenges in organizations, the following steps may be followed:

- Carefully state the company's AI policies [46].
- Establish a strong AI ethical committee.
- Collaborate with reliable stakeholders.
- Provide open-source toolkits to the AI trust pillars.
- Set up definable AI roles within your organization.
- Train and educate your AI professionals.
- Establish formal AI processes.
- AI literacy must be democratized across organizations.

AI is driving an enormous shift in business [31]. Adopting AI requires a strategic change that calls for a data-centric strategy and the inclusion of ethical measures to reduce the hazards involved [46]. To prosper in a dynamic world with increasing regulatory pressure and elevated customer awareness and expectations, organizations must prioritize quality management [34]. A combination of creativity and unwavering adherence to moral values is needed to establish long-term trust across the AI life cycle, from conception to commercialization.

Standards were created to satisfy the documentation requirements, performance monitoring, training requirements of staff, and regular updates on new requirements [38]. Documentation aims to maintain records, provide regular updates on documentation, and ensure version control. Standards are continuously evolving, so organizations should monitor for updates, and regular audits are recommended for compliance [33]. Some of the core organizational AI standards released by ISO and IEEE are shown in Table 1.

International Organization for Standardization (ISO) and International Electrotechnical Commission (IEC) form the specialized system for worldwide standardization. Following are some of the ISO standards for the organizations:

### 5.1.1. ISO/IEC 42001:2023

The first AI management system standard in the world, ISO/IEC 42001, offers helpful direction for this quickly evolving technological area. It tackles the particular difficulties posed by AI, including ethical concerns, transparency, and continuous growth. ISO/IEC 42001:2023 [53] is intended to gain credibility and establish a strong base for AIMS, or AI management systems. Benefits of using this standard are

- It helps to establish brand awareness and trust by showcasing RAI development and application.

- Maintain an edge and prepare the organization for the future by building the foundation for ethical AI practices.

- achieve compliance, control risks, improve productivity, and confidently traverse the constantly evolving regulatory environment by using an organized AI management system

It lays forth an organized strategy for businesses to balance innovation and governance while managing the risks and possibilities posed by AI.

### 5.1.2. ISO/IEC 23894:2023

The focus of ISO/IEC 23894 [54] is AI risk management. It recognizes that AI poses both new and distinct problems, even though it shares some concerns with conventional software systems (such as security flaws). These are derived from the main attributes of AI, which include its ability for data-driven learning, self-governing decision-making, and real-world interaction.

### 5.1.3. ISO/IEC 23053:2022

ISO/IEC 23053:2022 standard [55] provides a framework for AI Systems using machine learning (ML) and that describes an overall AI system. The framework explains the various system components and how they work within the AI ecosystem. Any organization that uses or implements AI systems is covered under this document, including government agencies, public enterprises, and non-profit organizations.

### 5.1.4. IEEE P2941.3

It is a standard [56] for representation and application Programming Interface(API) of Large scale Pre-trained AI Models . Large-scale pre-trained AI models, including their training, inference, transfer, mixture, instruct-tuning, representation, compression, distribution, and services, are supported by the standard's representation and API framework. The purpose of this standard is to make it easier to design and implement large-scale, pre-trained AI models across many platforms and domains.

### 5.1.5. IEEE 7000-2021

This standard [57] establishes a set of procedures that companies can use to include ethical ideals into concept exploration and development processes. It supports the elicitation and prioritization of ethical values by management and engineering in transparent communication with selected stakeholders. It involves the traceability of ethical values through an operational concept, value propositions, and value dispositions in the system design. It describes procedures that enable ethical values in a system of operations, ethical requirements, and ethical risk-based design to be traced back. This standard applies to companies of any size, each employing its own life cycle model.

The EU AI Act's organizational requirements for compliance have many characteristics with international standards like IEEE 7000 and ISO 42001. Creating a risk management system, keeping up with thorough technical documentation, and including human monitoring are a few of them. The broad procedures, guidelines, and practices that businesses put in place to control the creation, use, and utilization of AI technology are the main emphasis of these standards. Standards at the organizational level may cover a broad range of topics, such as:

1. Creating rules and regulations for the development and implementation of AI inside a company.

2. Bringing risk assessment and mitigation techniques into practice to handle the adverse impacts of AI systems [58].

3. Securing data utilization, algorithmic transparency, and explainability, as well as transparency and accountability in AI decision-making processes [59].

4. Throughout the AI lifecycle, data quality, privacy, security, and access should be managed through the establishment of data governance frameworks [60].

5. Educating employees on ethical AI practices, rules pertaining to AI, and responsible data practices [61].

### 5.2. Product Level AI Standards

Product level AI standards, on the other hand, concentrate particularly on the development, testing, and certification of individual AI systems.

i)Ensuring AI systems fulfill predetermined accuracy criteria and performance measures for the purposes for which they are designed. ii)Ensuring the reliability of AI systems under a range of circumstances and setting safety measures in place to prevent harm to individuals or the environment iii)Encouraging compliance with current standards and technologies as well as interoperability amongst various AI systems. iv)Including security mechanisms to protect against data breaches, illegal access, and invasions of privacy To assure the safety, quality, and compliance of AI systems with regulatory requirements, AI product-level standards are frequently more precise and technical, concentrating on the design and engineering components of AI systems and their use case. Their structure and logic are more akin to ISO 14971 for medical device risk management.

Product level standards focus on documentation requirements, Testing Procedures, Monitoring ,Maintenance, Regular updates on the standards , Compliance monitoring, Performance optimization , Security maintenance and User feedback integration. Some of the Product Level AI standards were released by ISO and IEEE are shown in the Table :2

### 5.2.1. ISO/IEC 27001: Information Security Management Systems

Risk of cyberattacks can be reduced, respond to evolving security risks, and ensure that assets like employee data, financial statements, intellectual property, and information entrusted to third parties remain undamaged, confidential, and available when needed by putting the information security framework outlined in the ISO/IEC 27001 standard into practice. Additionally, it offers a centrally managed framework that securely stores all information in one location. Throughout your business, prepare people, procedures, and technology to handle risks and other dangers related to technology. Protect data in all formats, including digital, cloud, and paper-based data; Save money by reducing expenses and increasing productivity.

### 5.2.2. ISO/IEC 27701: Privacy Information Management Systems

[https://www.iso.org/standard/71670.html] This document is an extension to ISO/IEC 27001 and ISO/IEC 27002 for privacy management within the context of the organization. It defines requirements and offers guidance for creating, implementing, maintaining, and continuously improving a Privacy Information Management System (PIMS).

### 5.2.3. ISO/IEC 29134: Guidelines for privacy impact assessment

[ Guidelines for privacy impact assessments and the format and content of PIA reports are provided in this document. It can be used to businesses of all shapes and sizes, including government agencies, non-profits, private and public enterprises. This document is relevant to everyone working on project design or implementation, including those running PII-processing data processing systems and services.

### 5.2.4. IEEE P7000 series: Ethically aligned design for AI Systems

Intelligent and autonomous technology systems are explicitly created to decrease the need for human intervention in our daily lives. By doing this, these new systems are also giving rise to worries about how they will affect people individually and as a society. Supporting beneficial effects, like process and resource optimization, better planning and decision-making, and finding valuable patterns in big data, are among the topics of current discourse. Concerns regarding probable invasions of privacy, discrimination, loss of skills, poor economic implications, threats to the security of vital infrastructure, and potential long-term detrimental effects on social well-being are all discussed and warned against.

### 5.2.5. NIST SP 800-53: Security and Privacy Controls for Federal Information Systems and Organizations

To safeguard organizational operations and resources, people, other organizations, and the country against a wide range of threats and risks-such as hostile attacks, human error, natural disasters, structural failures, foreign intelligence entities, and privacy risks. This publication offers a catalog of privacy and security measures for information systems and organizations. The measures are applied as a part of an organization-wide risk management strategy and are adaptable and adjustable. A variety of requirements arising from laws, executive orders, directives, rules, policies, standards, and guidelines, as well as mission and business needs, are addressed. Ultimately, security and privacy are covered by the consolidated control catalog from two perspectives: functionality such as., the effectiveness of the processes and functions offered and assurance, the degree of confidence in the security or privacy capability offered by the controls. Information technology products and the systems that depend on them can be made adequately trustworthy by taking functionality and assurance into consideration.

## 5.3. Global standards

The research, implementation, and application of AI systems around the world are governed by a complex network of recognized frameworks, rules, and regulations known as global AI standards. Important international organizations like ISO/IEC, which offers fundamental frameworks like ISO/IEC 23053 for machine learning systems and ISO/IEC 22989 for AI terminology, serve as the foundation for these standards. Regional initiatives also influence the landscape; the European Union's AI Act, for example, sets the standard for complete regulation, while national authorities in Asia and the United States, such as NIST, provide their own frameworks and standards. Together, these standards include important topics like governance frameworks, risk management, ethical issues, and technical requirements. They include specifications for system architecture, performance metrics, privacy, fairness, transparency, and data quality. While national laws like China's GB/T 40338 and Singapore's AI Governance Framework give more specialized advice, major frameworks like the IEEE standards and the OECD AI Principles offer global standards for the creation of ethical AI. These guidelines work together to provide a multifaceted strategy that guarantees AI systems are created and used responsibly, taking into account safety, ethics, and the influence on society. Some of the Global AI standards released by various organizations shown in the Table : 3

### 5.3.1. The OECD Principles on AI

The first intergovernmental AI standard is the OECD AI Principles. They support cutting-edge, reliable AI that upholds democratic principles and human rights. These guidelines, which were adopted in 2019 and revised in 2024, consist of five principles based on values and five proposals that offer policymakers and AI actors adaptable and useful direction.

### 5.3.2. The EU Ethics Guidelines for Trustworthy AI

According to the Guidelines, AI systems must satisfy seven essential criteria in order to be considered reliable, they are Human agency and oversight, Technical Robustness and safety, Privacy and data governance, Transparency, Diversity, non-discrimination and fairness and Accountability.

Table 1: Global AI Standards

| Standard/Framework | Organization | Key Components |
|---|---|---|
| ISO/IEC 23053 | ISO/IEC | Framework for AI Systems Using Machine Learning |
| ISO/IEC 22989 | ISO/IEC | AI Concepts and Terminology |
| ISO/IEC 38507 | ISO/IEC | Governance Implications of AI |
| IEEE 7001-2021 | IEEE | Transparency of Autonomous Systems |
| AI Act | European Union | Risk-Based AI Regulation |
| OECD AI Principles | OECD | Ethical AI Guidelines |
| GB/T 40338 | China | AI Risk Management |
| NIST AI 100-1 | NIST | AI Risk Management Framework |
| AI Governance Framework | Singapore | AI Deployment Guidelines |
| P 7009 | IEEE | Fail-Safe Design for AI Systems |
| AI Procurement Guidelines | WEF | AI Acquisition Standards |

*5.3.3. The IEEE Global Initiative for Ethical Considerations in AI and Autonomous Systems*

The IEEE Global Initiative aims to reframe the concept of success so that human progress can include the deliberate prioritization of individual, community, and societal ethical values. In order to promote these technologies for the benefit of humanity, The IEEE Global Initiative seeks to empower, educate, and train all stakeholders involved in the design and development of autonomous and intelligent systems to prioritize ethical issues. It also seeks to incorporate ethical aspects of human well-being that may not automatically be taken into account in the current design and manufacture of A/IS technologies

*5.3.4. The World Economic Forum's AI Governance Framework*

The main workstreams of the Alliance are focused on important areas of AI, encouraging ethical use and innovation. These workstreams concentrate on creating strong regulatory frameworks via resilient governance and regulation, integrating AI technologies across industries responsibly, and improving technical standards for sophisticated and secure AI systems.

*5.3.5. The Global Partnership on Artificial Intelligence (GPAI) Principles*

As a global reference point for particular AI issues, GPAI hope to reduce duplication, foster international collaboration, and foster trust in and adoption of trustworthy AI by offering a platform for exchanging multidisciplinary research and identifying critical issues among AI practitioners. The goal of all GPAI initiatives is to promote the ethical development of AI based on these values of inclusiveness, diversity, human rights, innovation, and economic prosperity.
.

*5.3.6. The AI Global Governance Commission's AI Ethics Guidelines*

In November 2021, UNESCO published the "Recommendation on the Ethics of Artificial Intelligence," which is the first global guideline on AI ethics and its been adopted by 193 Member States. Ten guiding principles define an approach to AI ethics that is centered on human rights, they are Proportionality and Do No Harm, Safety and Security, Right to Privacy and Data Protection, Multi-stakeholder and Adaptive Governance & Collaboration, Responsibility and Accountability, Transparency and Explainability, Human Oversight and Determination, Sustainability, Awareness & Literacy, Fairness and Non-Discrimination

*5.3.7. The Montreal Declaration for RAI*

Three primary goals comprise the Montréal Declaration for RAI development: 1. Create an ethical foundation for the creation and application of AI; 2. Direct the digital shift so that the advantages of this revolutionary technology are shared by all; 3. Establish a national and worldwide discussion forum to work towards a shared goal of fair, inclusive, and environmentally sustainable AI development.

*5.3.8. The AI4People Global Summit Declaration on Ethics and AI*

The work of the AI4People-Automotive Committee, which was formed to provide more detailed advice on particular ethical problems arising from autonomous vehicles (AVs), is presented in this document. Across all topic areas-human agency and oversight, technical robustness and safety, privacy and data governance, transparency, diversity, non-discrimination and fairness, societal and environmental wellbeing, as well as accountability-practical recommendations are offered for the automotive industry. This allows this research to differentiate between industry suggestions that create guidelines for businesses throughout the supply chain and policy recommendations that strive to help lawmakers set acceptable criteria. These suggestions may be used in the future by the automotive industry to decide on appropriate next actions and guarantee that autonomous vehicles abide by moral standards.

*5.4. Forthcoming AI standards*

In response to the rapid advancement and application of AI technology, the field of upcoming AI standards is changing quickly as of early 2024. A number of significant international organizations and regulatory agencies are attempting to create all-encompassing frameworks that will influence the advancement and application of AI in the future. An innovative legislative framework that paved the way for risk-based categorization and supervision of AI systems is the European Union's AI Act. In the meanwhile, organizations like IEEE and ISO are creating technical standards like the IEEE 7000 series, which focuses on ethical issues in AI design, and ISO/IEC 42001 for AI management systems. Guidelines for AI safety, risk management, and other topics are also being developed by national standards organizations, such as BSI in the UK and NIST in the US. The standards, many of which include adaptable frameworks that can adjust to the quickly evolving technological landscape and new difficulties in AI development, are anticipated to be implemented in phases over the course of 2024 - 2025. Some of the key aspects of these frameworks are Risk Management, Technical Requirements, Ethical Considerations and Governance requirements. Some of the forthcoming AI standards which are expected to be released by ISO and IEEE are shown in the Table : 4

*5.4.1. ISO/IEC 42005*

AI system impact assessment standard expected to be published in early 2025 . Organizations conducting AI system impact assessments for people and societies that may be impacted by AI systems and their planned and anticipated uses can find guidance in this publication. It covers guidelines for AI system impact assessment documentation as well as how, when, and at what stage of the AI system lifecycle to conduct such evaluations. This guidance also explains how an organization's AI risk management and AI management system can incorporate the AI system impact assessment methodology. This paper is meant for use by companies who create, supply, or use AI systems. Any business, regardless of size, type, or nature, can use this document.

*5.4.2. CEN-CENELEC*

In parallel, CEN-CENELEC is planning to provide the first iteration of the harmonized EU AI Act standards, which must be completed and available for public discussion by December 2024. CEN-CENELEC must ensure that EU standards and standardization deliverables comply with EU law on fundamental rights and data protection by submitting a Final Report to the European Commission by April 30, 2025.

*5.4.3. Final EU AIA*

Classification of AI systems as high-risk and prohibited AI practices is expected to be published on April 2025. A horizontal layer of protection, including a high-risk classification, is provided by the compromise agreement to make sure that AI systems that are not likely to seriously violate basic rights or pose other major hazards are not detected. AI systems which cause a small risk would be bound by very minimal transparency requirements, such as stating that the content was created using AI so that consumers can decide whether or not to utilize it further. Various high-risk AI systems would be approved, but only after meeting certain conditions and fulfilling specific obligations to be allowed to enter the EU market.

## 6. APPLICATIONS OF RAI

RAI applications focus on developing and deploying AI systems that are ethical, transparent, and beneficial to society. These applications prioritize fairness, accountability, and privacy in their design and implementation. Examples include:

- It acts as a tool for detecting and mitigating bias in hiring
- In healthcare, interpretable AI models for clear diagnosis
- Data analytics using machine learning techniques that protect privacy
- Autonomous car AI decision-making systems that uphold ethical standards
- Using AI to monitor the environment for sustainable resource management
- Financial services fair lending algorithms
- Organizations using AI ethics review boards to supervise AI initiatives

RAI aims to maximize the benefits of AI while minimizing potential harm, ensuring that AI systems align with human values and societal norms. Here are some of the applications of RAI

- **Healthcare:** RAI can be used in healthcare to improve patient outcomes, assist in diagnosis, and personalize treatment plans.
- **Finance and Economics:** RAI can be used in the finance industry to detect fraud, assess credit risk, and optimize investment strategies.
- **Agriculture:** RAI facilitates sustainable farming practices, waste reduction, and increased crop yields.
- **Education:** RAI can be used in education to personalize learning experiences, provide feedback to students, and assist teachers in lesson planning.
- **Transportation:** RAI can be used in transportation to optimize traffic flow, improve safety, route optimization, Autonomous vehicles and reduce emissions.
- **Sustainability ESG- Environment, social, and governance:** RAI can be used in Environment monitoring for carbon footprint analysis, resource management, and Bio-Diversity.
- **Industry:** RAI can be used in Industry for Quality Control, resource optimization, Risk Management, and smart manufacturing

Table 2: Forthcoming AI Standards

| Standard | Organization | Key Focus Areas |
|---|---|---|
| ISO/IEC 42001 | ISO/IEC | AI Management Systems Requirements |
| IEEE 7000 Series | IEEE | Ethical considerations in AI system design |
| IEEE 2089 | IEEE | Age-appropriate design for AI systems |
| EU AI Act | European Union | Risk-based AI regulation framework |
| ISO/IEC 42002 | ISO/IEC | AI Management System Guidelines |
| CEN/CENELEC AI Standards | European Standards Organizations | AI standardisation |
| AI Risk Management Framework (AI RMF) | NIST | Risk management and trustworthy AI |
| ISO/IEC 23894 | ISO/IEC | AI Risk Management Guidelines |
| IEEE 2863 | IEEE | Organizational AI Governance |
| ISO/IEC 42001 | ISO/IEC | AI Management System Requirements |
| AI Safety Standards | BSI (UK) | Safety and reliability of AI systems |

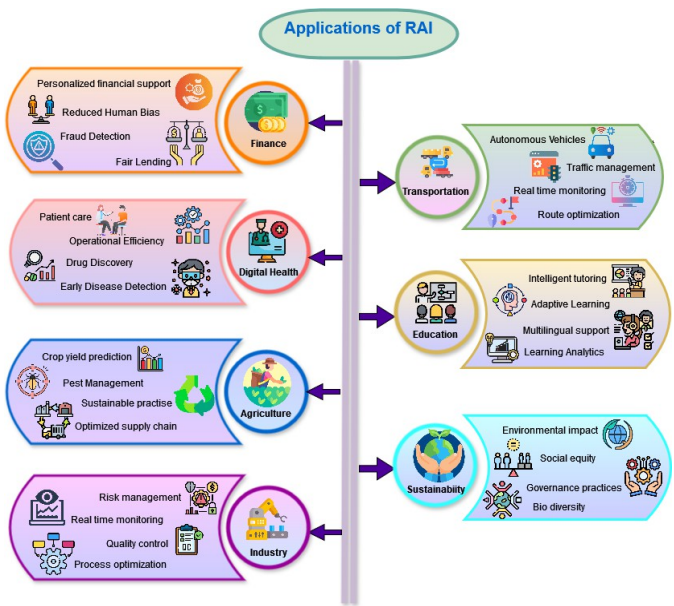

Figure 4: Applications of RAI

### 6.1. RAI in Digital Health

"RAI in digital health" refers to the morally and ethically sound deployment of AI technologies in the medical domain. It comprises creating and bringing into use AI systems that prioritize patient safety, privacy, and justice and that adhere to both ethical and legal constraints. The objective is to strike a compromise between innovation and the protection of human rights and values, all the while boosting diagnosis accuracy, expediting the healthcare system, and improving patient outcomes. It can speed up scientific advancements and discoveries that enhance human well-being in the healthcare sector. AI, for instance, has the potential to impact patient diagnosis, drug development, and healthcare delivery. Though the application of RAI in healthcare is still in its early stages, 94% of IT executives think RAI development deserve more attention.

AI in digital health makes use of these technologies to enhance patient outcomes, medical research, and healthcare delivery. It includes applications in drug development, disease diagnosis, therapy planning, and personalized medicine. Major issues of using AI in digital health are identity theft, lack of transparency, privacy invasion, and algorithm misuse arising from the gathering and use of personal data by AI and analytical algorithms [62, 63]. According to a survey conducted in the UK, 63% of respondents feel uneasy about AI taking over from doctors and nurses for certain jobs, such as making suggestions for treatment, and 49% say they would never give up their personal health information to create algorithms that could potentially enhance the quality of care [64]. The authors in [65] discussed the drawbacks and RAI solutions by considering the works published by various authors. Patients may not feel comfortable disclosing the data, and privacy issues result from managing sensitive health data. A possibility that bias in AI algorithms will result in inequities in healthcare. AI decision-making procedures are not transparent. There are difficulties in incorporating AI systems into the current healthcare framework by ignoring the knowledge of human and relying too much on AI. Suppose these ethical quandaries and issues are not sufficiently resolved when using AI for medical analytics and digital health, it may have a detrimental effect on patients as well as damage the standing of healthcare institutions [66]. Many nations have enacted data protection laws in response to these ethical quandaries. One such law is the UK's Data Protection Act 2018, which complies with the General Data Protection Regulation (GDPR) created by the European Union. A scientific and practical focus on the secure utilization of AI is intended by these regulations, which seek to increase people's confidence in sharing personal information with healthcare organizations. RAI ensures that AI algorithms are developed and deployed in a way that prioritizes patient safety, privacy, and autonomy. Some key principles of RAI in digital health include:

- **Transparency:** Information on the creation, training, and application of AI algorithms in clinical settings should be openly shared by healthcare providers and AI developers. This involves revealing the sources of the data that were used to train the algorithms as well as any potential drawbacks or biases in the system. Transparency is key to this. Training data sources, including genomic databases, medical imaging libraries, and electronic health records (EHRs). Manufacturers are required by the FDA's Software Precertification Program, for example, to submit comprehensive data on the creation, testing, and validation of AI algorithms. An integrated approach is required for RAI solutions in digital health. To protect patient information, this involves setting strong data protection and anonymization procedures into place. Transparent decision-making procedures are made possible by explainable AI models, and diverse and sample datasets have been developed to reduce bias. Accountability in AI healthcare applications is ensured by the establishment of ethical principles and regulatory frameworks. Moreover, encouraging human-AI collaboration improves healthcare results as opposed to displacing human experts. A regular auditing and effect assessment process for AI systems helps find possible problems. In the end, universal healthcare is advanced by guaranteeing fair access to AI-powered healthcare solutions. In order to solve current issues with AI technology research and implementation, RAI in Digital Health places a high priority on ethical considerations, transparency, and patient welfare. The application of AI in healthcare should be subject to clear lines of accountability, which should include procedures for handling mistakes, biases, and any other potential ethical issues.

- **Security and privacy of patient information:** The security and privacy of patient data are given the highest priority in the integration of RAI in healthcare, guaranteeing private and secure health data. Electronic health records (EHRs) are protected from unwanted access by access controls, secure data sharing procedures, and AI-driven encryption techniques. Audit trails keep track of data access and updates, and sophisticated machine learning algorithms identify any irregularities and breaches. AI methods that protect privacy allow analysis without jeopardizing patient identities, and open decision-making procedures uphold patients' autonomy and confidence. Moreover, AI-powered solutions adhere to strict data protection standards by complying with legal frameworks like GDPR and HIPAA. Healthcare providers may preserve the highest levels of data privacy and security, stop data breaches, and safeguard sensitive patient information by utilizing RAI.

*6.2. RAI in finance and economics*

In financial services and economics, RAI seeks to maximize its potential while minimizing risks, guaranteeing equity, and preserving the integrity and stability of financial systems. To do this, a number of important tactics are used. To avoid biases and guarantee fair results, machine learning algorithms that are cognizant of fairness are employed. In order to provide accountability and openness in financial decision-making, explainable AI models are built. To protect sensitive financial data, strong cybersecurity measures are strengthened and vulnerability evaluations are carried out on a regular basis. The application of AI in financial markets is governed by regulatory frameworks that provide supervision and compliance. Through focused programs, finance professionals are retrained and upskilled to work with AI systems more efficiently. Regular audits and effect assessments of AI systems are conducted, and strict privacy and data protection regulations are implemented.

For the organizations to discover, evaluate, and reduce possible risks while maintaining regulatory compliance and moral decision-making, RAI, or AI, is essential. Financial institutions can improve their risk management skills by utilizing RAI techniques like machine learning, natural language processing, and predictive analytics. These techniques can help identify early warning indications of credit risk, market risk, and operational risk. Real-time monitoring, automatic risk assessment, and data-driven insights are all made possible by RAI-powered systems, minimizing possible losses and enabling proactive risk mitigation. Furthermore, RAI promotes trust and confidence in financial institutions by guaranteeing accountability, explainability, and openness in AI-driven decision-making. Financial firms can enhance regulatory compliance, preserve the stability and integrity of the financial markets, and maximize risk-reward tradeoffs by incorporating RAI into their risk management frameworks.

Organizations that demonstrated the developing practices were starting to introduce new procedures and structures or modify already-existing ones, yet certain emerging structure such as inflexible organizational incentives in situations with high levels of inertia may impede rather than promote RAI research. In particular, the remaining emerging work practices lessened the labor burden on individuals by identifying organizational processes and policies that support their RAI work and by managing the transitions involved in adopting those new work practices. This helped to enable RAI work better. This frees up time that people can use to concentrate on actual RAI work [67]. People who are concerned about algorithmic responsibility issues can easily dedicate their time and effort to making progress on the specific issues within their functions because, in the ideal future, organizational structures and processes will fully provide mechanisms for monitoring and adapting system-level practices to incorporate and address emerging ethical concerns. Full-time positions in internal advocacy and change management would be awarded to individuals who possess the necessary abilities, training, and motivation to concentrate on that work. These individuals would also be able to provide guidance and support to other individuals as they collaborate to bring about systemic change both within and outside of their organizations. Then, those developing RAI may concentrate on their particular profession instead of altering the workplace to make it RAI in finance and economics refers to the ethical and transparent use of AI technologies in the financial industry. This includes ensuring that AI algorithms are fair, unbiased, and accountable, and that they do not perpetuate discrimination or harm to individuals or society. Some key principles of RAI in finance and economics include:

- **Fairness:** AI systems ought to be created with the intention of treating every person equally and without prejudice, irrespective of their socioeconomic background, gender, or race.

- **Transparency:** In order for users to comprehend how decisions are made and contest them if necessary, the decision-making process of AI algorithms should be clear and explicable.

- **Accountability:** Businesses utilizing AI in economics and finance ought to answer for the results of their algorithms.

*6.3. RAI in Agriculture*

RAI in agriculture refers to the ethical and sustainable use of AI technologies in farming practices. This includes ensuring that AI systems are used in a way that benefits both farmers and the environment, while also taking into account social and ethical considerations.

A clear understanding of the ethical framework can help us build and develop AI-based tools for digital agriculture. Developing ethical technology helps win the trust of stakeholders in the agricultural industry, especially farmers, and ensures ethical validity and accountability [68]. Social, ethical, technological, and political factors were all recommended for agriculture technology providers and politicians. ATPs are urged to evaluate the ethical aspects of developing AI systems through risk assessments and to take precautionary measures and preventative measures into account. Along with working together with different stakeholders in the value chain, these businesses should address any ethical concerns that may arise from the use of these technologies. Only a few of the best practices and procedures that were suggested in this research were raising awareness, offering farmers training opportunities, HCAI, interpretability and usability, transparency, and technological features that facilitate data transfer. Policymakers were advised to develop new regulations or amend current ones pertaining to farm infrastructure and data, finance research and development projects, back initiatives for living labs, and offer instructional programs to farmers and other participants in the smart farming system. Authors in [69] grouped the results into four themes: data access and associated ethical issues; laws and how they affect research on AI food system technologies; obstacles to the creation and uptake of AI-based food system technologies; and trust bridges that researchers believe are crucial to overcoming the obstacles they identified. Agricultural systems are under distress due to factors including increasing laws and deteriorating environmental conditions, which are creating opportunities for technological solutions. They highlighted the dysfunctional nature of the technology development and adoption process, which raises the possibility of these windows closing early. The knowledge gained from these interviews can assist organizations and governments in creating laws that will maintain open borders by fostering the development of technologies deserving of the labels "responsible" or "trustworthy" AI and bridging gaps between competing interests.

Some key principles of RAI in agriculture include:

- **Transparency:** Farmers need to be fully aware of how AI technologies are being applied to their operations, including the collection, analysis, and use of data in decision-making.

- **Accountability:** Businesses and institutions creating AI-based agricultural technology ought to answer for how their goods affect farmers, customers, and the environment.

- **Privacy and data security:** In accordance with data protection laws, farmers' data should be safeguarded and utilized responsibly.

*6.4. RAI in Education*

RAI in higher education refers to the ethical and transparent use of AI technologies in academic settings. This includes ensuring that AI systems are designed and implemented in a way that respects privacy, fairness, accountability, and transparency.

Some key principles of RAI in higher education include:

- **Fairness:** Ensuring that AI systems do not discriminate against individuals based on factors such as race, gender, or socioeconomic status.

- **Transparency:** Ensuring that teachers and students are aware of how their data is being utilized and giving clear explanations of how AI systems make decisions.

- **Privacy:** Making sure that personal information is only utilized for the purposes for which it was collected.

- **Accountability:** Holding organizations and people accountable for AI's decisions

The Artificial Intelligence High-Level Expert Group (AI HLEG) acknowledges that "AI systems offer great opportunities, but they additionally create certain risks that must be managed appropriately and proportionately." Thus, in order to advance technology, it is critical to invest in education as well as to ensure that the next generation of experts can create technology in a way that preserves European values. In addition to teaching young people with advanced skills how to program, Higher Education (HE) should also teach all students how to comprehend the consequences of AI and how to influence its ethical use. While AI systems have many potential, they also pose concerns and may violate democratic or ethical norms pertaining to individual responsibility, inclusiveness, privacy and transparency. Higher Education (HE) significantly contributes to the development of innovative, secure, and ethical AI. Authors in this study recommended the following suggestion, using national education policies, coordinate the integration of trustworthy AI into courses to guarantee a consistent adoption, encourage higher education institutions to hire specialists and invest in teacher training to acquire the necessary knowledge to instruct in Trustworthy AI and encourage the use of multidisciplinary teams in the classroom by giving them credit and giving it a high priority in the curriculum [70].

The International Society of Artificial Intelligence in Education's work defines the field of AIED, founded on genuine objectives to identify, develop, and implement the finest teaching and learning strategies for as many students as possible [71]. The AIED Community should be aware that hazards to persons and groups as well as other forms of prejudice can be encoded by data and algorithms, and that well-intentioned ideas developed in research labs frequently do not translate to mainstream applications. Recognizing the possible drawbacks and abuses of AIED research won't make its goals or objectives less relevant; rather, it will likely make it more valuable and contribute to the development of new AIED techniques and technologies; Even though the goals of AIED have always been clearly linked to frontline education, the field's understanding of the larger implications of AI for learning, human functioning, and sociocultural effect is still somewhat restricted. The AIED Community is beginning to recognize the importance of a broader agenda on ethical AI, as seen by a number of recent papers on the Ethics of AI in Education [72, 73].

*6.5. RAI in Transportation*

RAI in transportation refers to the ethical and safe use of AI technologies in the design, development, and operation of transportation systems. This includes ensuring that AI algorithms are transparent, fair, and accountable, and that they prioritize safety, efficiency, and sustainability.

Some key principles of RAI in transportation include:

- **Safety:** Passengers, pedestrians, and other road users' safety should be the first priority when designing AI systems. This entails making certain that AI algorithms are strong and dependable as well as capable of prioritizing safety when making decisions in the moment.

- **Transparency:** In order for users to comprehend how decisions are made and have faith in the dependability of the system, AI algorithms utilized in transportation should be clear and understandable.

AI ethical guidelines are now being developed by a number of organizations and institutions [74] such as [EU, 2019] and standards [1S0 2021], in addition to generic ethical standards for system designs such as IEEE 7000 [IEEE 2021], Engineering ethics and machine ethics are two separate ethical challenges that need to be addressed with regard to AI-based safety-critical systems[TUVR. 2022]. Authors in [75] study compiles and organizes a large body of disorganized literature on the use of AI to the development of safety-critical systems for the transportation and industrial sectors, ranging from next-generation autonomous systems to conventional functional safety. Automated decision-making instantiations of a particular kind have already been employed in conjunction with compensatory measures (like safety bags) in the development and certification of automatic safety-critical systems like railway interlocking. Furthermore, the most recent Advanced Driver Assistance Systems frequently use AI technology to construct certain heteronomous safety roles that call for human supervision. Overall, because of the potential benefits to society and the general interest of industry, a clear path is to be laid for the development, certification, and assessment of AI-based safety-critical systems. Therefore, authors anticipate that the multidisciplinary domains of safety-critical systems, trustworthiness, and AI research will continue to be dynamic and robust in the years to come. The intricacy of the smart transportation system(STS) is increased by environmental pollutants, excessive traffic, and unpredictable congestion. The future STS will be able to address each of these issues going forward. As part of the transportation services, it opens the door to improving user security, $CO_2$ emission control, congestion control, etc. The Intelligent Transport System incorporates a variety of AI solutions due to the growing population and individual transportation needs. The work in [76] focused on a variety of AI for future smart transport management systems. The authors discussed the following areas: the issues with the current transportation system, various AI technologies related to smart transportation, and emerging AI technologies that manage future smart transportation services.

The authors in [77] presented a thorough framework for vehicle safety designed especially for smart cities. To improve road safety, their framework smoothly incorporates a range of sensors, such as alcohol, ultrasonic, and eye blink sensors. In order to improve safety on smart city roads, eye blink sensors are used to quickly identify possible risks and notify drivers through aural cues. Additionally, ultrasonic sensors enable more efficient traffic flow by providing real-time information on the speeds of nearby vehicles. It includes a unique sensor that efficiently monitors the driver's alcohol levels in order to address concerns over alcohol intake and its possible effects on road safety. The system automatically modifies the dosage when there is a high alcohol concentration using GPS and GSM technology.

*6.6. RAI in Sustainability ESG- Environment, social and governance*

RAI on sustainability, environment, social, and governance (ESG) refers to the use of AI technology in a way that prioritizes ethical considerations and promotes positive outcomes for the planet and society. This includes ensuring that AI systems are designed and deployed in a way that minimizes their environmental impact, respects human rights and promotes social equity, and upholds good governance practices.

Some key principles for RAI on ESG include:

- **Transparency:** To enable accountability and supervision, AI systems should be clear about their decision-making procedures and results.

- **Fairness:** In order to prevent prejudice and discrimination, AI systems should be developed and implemented in a way that supports equity and justice.

- **Data security and privacy:** AI systems must respect individual.

Integrating RAI within Environmental, Social, and Governance (ESG) frameworks is crucial for the ethical and sustainable deployment of AI as technology becomes more and more integrated into commercial processes. This study looks at how top businesses match their ESG objectives with RAI. Interviews with 28 prominent figures in the business revealed a close connection between RAI and ESG practices. Nonetheless, there is a notable discrepancy between public disclosures and internal RAI standards, underscoring the necessity of stronger governance, increased board-level experience, and staff involvement. Our main suggestions for bolstering RAI initiatives center on openness, interdisciplinary cooperation, and smooth integration with current ESG frameworks [78]. A unique ESG-AI framework was presented by the authors. It consists of three essential elements and was created using insights from interactions with 28 businesses. The framework, which was created in conjunction with industry practitioners, offers an organized method for this integration and three essential elements are RAI deep dive assessment, AI governance indicators, and AI use case [67].

Investors and other users can evaluate the materiality of AI use with the use of the ESG-AI framework, which gives a summary of the social and environmental effects of AI applications. Additionally, it makes it possible for investors to assess a company's dedication to RAI through well-organized interactions and in-depth analysis of particular risk areas.[78]. The authors in [78] introduced the reliable and statistically valid RAI measurement tool that businesses may use to gauge the level of RAI maturity in their AI capabilities. According to the "referenced ranking score" derived from the examination of RAI principles, the study included the following categories: explainability, fairness, data management, and security. SDG 2 is included into Ecuador's agrarian policy, which promotes the productivist model with a focus on economic development and minimally assumes the care of agrobiodiversity. This approach does not, in fact, lead to the aim of ending hunger and malnutrition. Although SDG 12 is included in the policy, which emphasizes sustainable consumption and production, efficient use of natural resources, and the advancement of science and technology for sustainability, there are no guidelines to prevent food waste, waste management, or the need for businesses to adopt sustainable practices, indicating a lack of will to encourage a shift towards sustainable production and consumption [79].

In order to overcome the obstacles, businesses should strengthen board-level RAI expertise, make RAI policies more transparent, raise employee awareness and participation, and establish specific AI goals.

*6.7. RAI in Industry*

RAI in industry refers to the ethical and transparent use of AI technologies in various sectors. This includes ensuring that AI systems are developed and deployed in a way that prioritizes fairness, accountability, and transparency.

Some key principles of RAI in industry include:

- **Fairness:** Making sure AI systems don't discriminate against people or groups on the basis of racial, gender, or socioeconomic background, for example.

- **Accountability:** Ensuring that there are procedures in place to resolve any negative effects and holding AI system creators and users responsible for the decisions these systems make.

- **Transparency:** ensuring that users are aware of the constraints and prejudices of AI systems and giving lucid explanations of how these systems arrive at judgments.

- **Privacy:** protecting employee's information

To accomplish organizational and practitioner-level engagement and promote RAI learning, the authors in [80] suggested an innovative stakeholder-first instructional strategy that makes use of interactive case studies. Collaborate with Meta, a global technology business, to create and present RAI courses to a wide range of employees. Findings show that attendees thought the seminars were interesting and that the knowledge and desire to use RAI in their job were growing. RAI ought to take values into account. Key stakeholders who own, utilize, and are impacted by these systems should have their values and priorities taken into account by RAI. From the beginning, the system design should be guided by these values. A multi-vocal literature review's findings served as the foundation for the RAI Pattern Catalogue. Instead of remaining at the concept or algorithm level, concentrate on practices that stakeholders in AI systems can implement to guarantee that the created AI systems are accountable across the whole engineering and governance lifecycle. Multi-level governance patterns, trustworthy process patterns, and RAI-by-design product patterns are the three categories into which the RAI Pattern Catalogue divides the patterns. For stakeholders looking to apply RAI, these patterns offer methodical and practical direction.

AI systems must take into consideration their wide variety of stakeholders in order to be genuinely responsible. For them to have a significant influence and decision-making authority during the design, development, and deployment of AI systems, their involvement is crucial. This calls for a setting where a range of stakeholders can feel empowered to implement changes and realize their goals. To guarantee that the systems that are produced have a good and significant impact on the socio-technical space, RAI must consider human issues. This involves admitting that technologies are not neutral artifacts but rather have an impact on and are influenced by the communities in which they are found [9, 81].

## 7. RAI in Technology

*7.1. RAI in blockchain*

The importance of RAI in the context of blockchain technologies follows from the significance of blockchain systems' role in making critical decisions, managing sensitive information, and having huge impacts on financial infrastructures, supply chains, and many other aspects. The ethical, transparent, and accountable operation of AI systems that become part of blockchain applications becomes important in view of the decentralized nature. The domains of applications of blockchains vary from financial to healthcare, in which sensitive data processing and vital choices are made; the integration of AI models with blockchain technology demands fairness, transparency, and accountability to maintain the trust and reduce the risks. RAI effectively protects from unethical decision-making and violation of privacy and ensures that AI systems operate within the ethical limit.The role of RAI in Blockchain Applications are described below.

*7.1.1. Enhancing Trust and Transparency:*

Blockchain systems are intended to be transparent and tamper-proof, so whenever AI algorithms are implemented, they have to stick to these principles. RAI guarantees that the AI models used in blockchain deliver clear and understandable explanations for their choices, thereby increasing trust among each stakeholder. This is especially important in applications like financial transactions, where the outcomes of AI-driven decisions are very important. RAI improves trust and transparency in blockchain applications by providing explainable AI-driven processes, guaranteeing fair and bias-free decisions, and protecting user privacy. Some of the studies show how adopting RAI helps blockchain systems to establish credibility and accountability among a variety of industries, like finance and supply chain management. In blockchain systems like decentralized finance, AI algorithms often make decisions, such as loan approvals or risk assessments. By applying the RAI principles, these decisions are made transparent and explainable. Users can understand why certain actions are taken, increasing trust in the system. For example, platforms like Aave and Compound employ AI-driven models to manage liquidity pools [82]. Through RAI, these models ensure that stakeholders can see the rationale behind interest rates and borrowing limits. In Compound, a decentralized finance platform, AI models help determine interest rates based on supply and demand. By integrating RAI practices, they ensure that the AI models decisions on interest rates are transparent. Users can easily verify how rates are calculated, enhancing trust between borrowers and lenders. Smart contracts are self-executing agreements whose terms and conditions are kept directly within code on a blockchain. When AI is integrated with smart contracts, RAI guarantees all parties understand the contract's logic and decisions. For example, if AI start a smart contract in supply chain management considering external data (such as product delivery status). RAI ensures that these events are transparent, reducing disputes and misunderstandings. AI-powered smart contracts monitor shipping and product status on the blockchain platform, which is used to track product origins in supply chains. The integration of RAI in blockchain platform enables that all AI decisions like paying money upon delivery are transparent to customers, preventing disputes and increasing trust [83]. Blockchain technology is used in financial systems wherein fraud detection is critical. AI models analyze transaction patterns in order to detect fraud. RAI ensures that fraud detection systems are transparent and accountable. For example, users must be able know why certain transactions are identified as suspicious, allowing them to trust the system's integrity. Chainalysis, a blockchain intelligence company, employs AI models to monitor suspicious activity on various cryptocurrency exchanges [84]. By utilizing RAI concepts, their models are transparent, explaining to authorities and platform users why some transactions are identified as fraudulent. This builds confidence in the platform and helps to avoid legal or regulatory challenges. In decentralized systems, AI helps in administrative tasks like stakeholder voting and decision-making. RAI makes sure the AI algorithms used in voting systems are transparent and not manipulated. This builds trust in decentralized governance models because stakeholders verify the validity of decisions. AI helps with governance at MakerDAO, a decentralized organization that manages the stablecoin DAI, by analyzing ideas and voting patterns. MakerDAO uses RAI to ensure transparency in AI's role, allowing members to review and verify AI-driven recommendations, resulting in a better and more transparent decision-making method [85].

*7.1.2. Fairness and Bias Mitigation:*

AI in blockchain applications needs to be free of biases that could lead to unequal treatment of people or groups of people. AI is used in DeFi platforms to evaluate creditworthiness and make automated trading decisions [86]. RAI ensures that AI models are periodically evaluated and trained on diverse datasets that fairly describe all groups. By employing fairness constraints and ongoing bias detection, DeFi platforms can provide equal lending and borrowing opportunities to users from various backgrounds. AI is used to analyze patient data and recommend treatments in blockchain-enabled healthcare systems. However, if AI model is trained on data that ignores specific ethnic or gender groups, it makes unequal recommendations [87]. This bias can lead to inadequate care for vulnerable people. RAI ensures that healthcare AI models are trained on a wide range of patient datasets and are regularly tested for bias. Furthermore, these models are subjected to fairness evaluations to avoid systemic bias. RAI also ensures transparency, allowing patients and doctors to understand how decisions are made, resulting in more equitable treatment. Smart contracts, which are self-executing contracts on the blockchain, can be biased if the AI models that cause them use incomplete or distorted data. This bias may favor some parties over others, particularly in global supply chains or legal agreements. RAI promotes transparency in AI-powered smart contracts. RAI ensures fair and equitable contract execution by incorporating bias checks and fairness algorithms. Furthermore, RAI promotes transparency, allowing stakeholders to better understand the logic behind contract executions and resolve disputes as required. To detect suspicious activities in blockchain-based fraud detection systems, AI models analyze transaction patterns. However, if the AI model contains historical bias, these systems may unfairly target specific demographic groups. This could result in unfair consequences, such as blocking transactions for certain groups more often than others [88]. RAI reduces bias in fraud detection by conducting fairness audits and ensuring that AI models are constantly monitored for demographic imbalances. This not only prevents discrimination, but it also increases trust in the fraud detection system by ensuring that all users are treated fairly. Table. 3 represents how RAI ensures fairness and mitigates bias in blockchain applications.

Table 3: Fairness and Bias Mitigation in Blockchain through RAI.

| Ref.No. | Area | Challenge | RAI Solution | Example | Impact of Bias | RAI Methods Used |
|---|---|---|---|---|---|---|
| [86] | Decentralized Finance | AI models may unfairly assess creditworthiness, leading to unequal access to loans. | RAI ensures that AI algorithms are trained on diverse datasets and regularly audited for biases. | Aave ensures RAI principles prevent biases in lending assessments. | Bias could lead to financial exclusion for minority groups. | Bias detection algorithms, regular audits, diverse training data |
| [87] | Healthcare | AI models may introduce biases in diagnosis and treatment recommendations based on biased data. | RAI monitors healthcare AI predictions to ensure fair and unbiased medical treatment. | Blockchain-based AI ensures equitable healthcare recommendations for all individuals. | Bias may lead to inadequate treatment for underrepresented populations. | Data anonymization, fairness evaluations, continuous testing against diverse population samples |
| [88] | Smart Contracts | AI-triggered smart contracts could unintentionally favor certain parties over others globally. | RAI guarantees transparency in AI logic and prevents bias in contract execution. | Provenance uses RAI to ensure fair execution of smart contracts in global supply chains. | Biased contracts could disrupt fair trade, leading to unequal payment terms or conditions. | Algorithm transparency, fairness constraints in automated contract terms |
| [89] | Fraud Detection | AI models may disproportionately flag certain demographic groups for fraudulent activities. | RAI applies fairness checks and audits to ensure fraud detection systems are unbiased. | Chainalysis employs RAI to avoid disproportionate fraud flagging across demographics. | Bias could unfairly target minorities, leading to wrongful suspicion or blocked transactions. | Fairness constraints, continuous monitoring for demographic imbalances in flagged cases |
| [90] | Governance in Blockchain | Decision-making processes in decentralized systems may be biased if not monitored. | RAI ensures equitable participation in governance by monitoring and mitigating algorithmic biases. | In MakerDAO, RAI promotes fairness in decision-making by ensuring AI-driven proposals are unbiased. | Bias in governance could result in power concentration or exclusion of minority voices. | Algorithmic voting fairness, bias detection in AI-based decision-making |
| [91] | Supply Chain Management | AI-driven predictions in blockchain-based supply chains may favor certain suppliers or regions. | RAI ensures equal treatment of suppliers by eliminating bias in AI-driven decisions. | AI in VeChain blockchain ensures fair distribution of resources by preventing region-based biases. | Bias could lead to unfair supplier favoritism, creating regional economic inequalities. | Data diversification, algorithm transparency, fairness audits in prediction models |

### 7.1.3. Predictive Analytics:

Predictive analytics plays an important role in blockchain, as AI can analyze massive amounts of decentralized data to forecast patterns, detect potential security threats, and identify market opportunities. AI models analyze patterns in blockchain transactions, network traffic, and user behavior to make predictions which will be beneficial for industries such as finance, healthcare, and supply chain management. Although blockchain networks possess inherent security, they remain vulnerable to some particular cyber threats such as hacking, double-spending attacks, and fraudulent transactions. AI models trained on historical data have the ability to predict potential vulnerabilities by examining irregular transaction patterns or detecting new threats within the network. Through the use of RAI, these models have the ability to give priority in providing security measures and guarantee that the algorithms do not unfairly target specific users or entities based on biased datasets. Predictive analytics help businesses to identify new market opportunities by analyzing patterns in blockchain data, such as supply chain efficiency or product demand across global markets. Blockchain data provide real-time insights into supply chain activities. RAI ensures that predictions made from blockchain data are fair, transparent, and accountable. One of the most significant challenges in predictive analytics is the risk of bias, particularly when algorithms are trained on incomplete or distorted information. RAI ensures that AI models used in predictive analytics are trained on diverse, representative datasets, which reduces the possibility of biased results. This is critical in industries such as finance and healthcare, where biased predictions can result in unequal access to services.

### 7.2. RAI in Federated Learning

A federated learning system enhances information security and fosters RAI while allowing numerous clients to learn a general model without exchanging local data. Distributed storage systems containing sensitive data have made centralized Machine and Deep Learning (ML/DL) techniques more difficult to use .Federated Learning (FL), a collaborative and privacy-preserving ML/DL technique, has emerged to solve data privacy problems. Because local participant data is kept secret while a global, collaborative model is being developed, FL guarantees data privacy by design. However, as there is an increasing need for confidence in model predictions, data privacy and performance alone are insufficient. So the RAI strategies are introduced to ensure robustness, fairness, explainability, and accountability as crucial pillars. The development of solutions that can measure the trustworthiness level of FL models, as well as the identification of trustworthiness pillars and evaluation metrics specifically pertinent to FL models, require more investigation. RAI is playing a critical role in the development and deployment of federated learning in several key ways:

### 7.2.1. Enhancing Trust :

Developing AI systems that are transparent and explicable is becoming more and more important, particularly for high-stakes applications. Making AI decision-making procedures easier for stakeholders and users to interpret and comprehend is a requirement of RAI. RAI methods combined with federated learning can promote accountability and trust in the models that are used. New laws and policies like the GDPR, the OECD AI Principles, and UNESCO's Ethics of AI Recommendations are frequently in line with RAI principles. For federated learning systems to be widely adopted and implemented, adherence to these ethical and legal frameworks is essential. RAI governance, which includes public involvement and institutional supervision, can guarantee that federated learning is created and applied in a way that advances society. RAI highlights the necessity of developing dependable, resilient, and accountable AI systems. Frameworks such as the "FederatedTrust" taxonomy offer a thorough method for assessing federated learning models' reliability in a number of areas. Protecting individual privacy is a crucial ethical concern that needs to be addressed as AI grows more widespread. The necessity for moral AI systems that respect people's privacy and adhere to moral prin-ciples is examined in this article. Using an interdisciplinary approach, the study looks at cutting-edge algorithmic methods including federated learning, homomorphic encryption, differential privacy, international legal frameworks, and ethical standards. The study comes to the conclusion that these algorithms successfully improve privacy protection while striking a balance between the necessity to secure personal data and the usefulness of AI[ Radanliev,2024]. In order to harness the power of AI in a way that respects and protects individual privacy, the paper highlights the significance of a comprehensive approach that blends technological innovation with ethical and regulatory policies. The study also highlights the importance of federated learning in decentralizing the process of data analysis. Federated learning avoids the privacy hazards of data transit and central storage by training AI models on users' devices instead of centralizing data. By ensuring that learning takes place across several devices, this methodology updates the model without jeopardizing individual data points. All things considered, our study proves that these algorithms can be used strategically to protect privacy in AI implementations. In addition to answering the research issue, the incorporation of these technologies into AI systems represents a significant advancement in the field of ethical AI development. These results provide a roadmap for utilizing AI while maintaining the highest standards, opening the door for additional study and useful applications.

The legal and regulatory framework governing AI responsibility is always changing. GDPR has important ramifications for AI accountability in this respect (Cristianini and Scantamburlo, 2019; GDPR, 2023). The people in charge of developing and utilizing AI decision-making systems need to accept accountability for the decisions the system makes. This entails keeping an eye out for biases and mistakes in the system and taking the appropriate action to reduce any negative impacts.

To guarantee that personal data is not compromised, AI systems must also respect individual privacy and follow stringent data protection requirements. Eliminating discriminatory consequences in AI decision-making processes requires addressing bias in AI algorithms and placing a high priority on fairness. It is crucial to preserve transparency and accountability in AI decision-making. As a result, it's important to look at accountability in AI, from model developers to users. AI in Warfare and Defence also highlights the moral ramifications of AI-driven decision-making in military applications [Dasgupta et al., 2020]. In particular, the issues of autonomous weapons and moral responsibility are examined.

Due to the existence of dispersed data silos that include sensitive information, the Internet of Things' (IoT) and Edge Computing's explosive growth has created difficulties for centralized machine learning and deep learning (ML/DL) techniques. Federated Learning (FL), a collaborative and privacy-preserving ML/DL technique, has emerged to solve data privacy problems. Because local participant data is kept secret while a global, collaborative model is being developed, FL guarantees data privacy by design. However, as there is an increasing need for confidence in model predictions, data privacy and performance alone are insufficient. Robustness, fairness, explainability, and accountability have been identified as key pillars in the literature's many approaches to trustworthy ML/DL (apart from data privacy). However, more study is needed to determine the evaluation metrics and trustworthiness pillars that are particularly pertinent to FL models, as well as to create methods that can calculate the degree of trustworthiness of FL models. In addition to introducing a thorough taxonomy with six pillars (privacy, robustness, fairness, explainability, accountability, and federation) and more than 30 metrics for determining the trustworthiness of FL models, this work looks at the current standards for assessing trustworthiness in FL. The trustworthiness score of FL models is then calculated using an algorithm called FederatedTrust, which is based on the metrics and pillars found in the taxonomy. FederatedScope is a well-known FL framework, and its learning process include a FederatedTrust prototype. Lastly, to show the usefulness of FederatedTrust in determining the reliability of FL models, the authors conducted five experiments with various FederatedScope setups (with varying participants, selection rates, training rounds, and differential privacy). Taking into account a real-world IoT security use case, two experiments make use of the N-BaIoT dataset and three experiments use the FEMNIST dataset [ SÃ¡nchez,2024]. The development of a new taxonomy that computes the reliability of FL models using the most pertinent pillars, concepts, and metrics. Important factors used to assess the reliability of federated and conventional ML/DL models were examined, contrasted, and assessed in order to develop such a taxonomy. More specifically, the taxonomy's primary building pieces were determined to be the following six pillars and over 30 metrics: federation, explainability, accountability, robustness, privacy, and fairness. Every pillar offers new measurements in contrast to previous research. Creating accountability systems and trustworthiness metrics promotes trust in the use of federated learning. Significant concerns about their responsible usage have been brought up by the integration of federated learning (FL) and AI in smart mobility.For intelligent transportation systems to be stable and sustainable, RAI is essential. Research on the proper use of FL and AI in this field is still in its early stages, despite its significance, and there are few comprehensive studies examining how they interact. Examining how FL functions in smart transportation and how RAI influences dispersed smart transportation. Finally, they go over the difficulties in creating and executing responsible FL in smart transportation and offer some possible fixes. Higher levels of intelligence, personalization, safety, and transparency are anticipated in intelligent transportation systems through the integration of RAI and federated learning[Huang,2024].

### 7.2.2. Fairness and Bias Mitigation:

A decentralized method for machine learning, federated learning (FL) allows several clients to work together on model training while protecting data privacy. FL is not impervious to issues of bias and fairness, though. Techniques like regularization, data augmentation, and fairness metrics can be used to lessen these problems. To lessen bias in the global model, FL frameworks can also be made to include fairness-aware aggregation techniques like weighted averaging or clustering-based strategies. Organizations may create more inclusive and equitable AI models that foster stakeholder trust and transparency by giving fairness and bias mitigation top priority in FL. A decentralized method for machine learning, federated learning (FL) allows several clients to work together on model training while preserving data privacy. Concern over the creation and use of RAI technology is growing as a result of the quick growth of AI applications and technologies. Massive volumes of data, sometimes including private and sensitive user information, must be gathered from various locations or nations in order to build AI technologies or machine-learning models. A brute force approach to the collection and integration of this data is prohibited by privacy, security, and data governance limitations. Thus, maintaining user privacy while developing high-performance models is a significant difficulty. Authors examined how federated learning has recently advanced in tackling this issue within the framework of computing that protects privacy. Federated learning offers strong incentive systems, high security and privacy guarantees, and the ability to train and employ global AI models across numerous decentralized data sources. The history, goals, definitions, architectures, and uses of federated learning as a novel approach to creating responsible, privacy-preserving AI ecosystems are covered in this article[ Yang, Q. (2021)].

*7.2.3. Data Privacy :*

Federated Learning (FL), which allows several clients to jointly train machine learning models without disclosing their local data, offers a strong framework for safeguarding data privacy. FL lowers the risk of data breaches and unwanted access by keeping client data on-premises and decentralized. Additionally, FL uses methods like homomorphic encryption, safe multi-party computation, and differential privacy to guarantee that only aggregated model updatesâ€"not raw client dataâ€"are exchanged. FL enables enterprises to get insights from dispersed data while protecting the security and integrity of sensitive data by employing cutting-edge cryptographic algorithms and retaining data locality.

Improving information security and advancing RAI, federated learning allows numerous clients to train a general model without sharing local data. However, weight divergence occurs, particularly for complicated graph data extraction, because the data of various clients in the system are not independently and identically distributed (IID). In order to increase the federated learning system's resilience in graph data, this paper suggests a novel feature-contrastive graph federated (FcgFed) learning technique. In order to examine graph information, we first create an architecture for FcgFed learning systems. In order to mitigate the weight divergence in federated learning, we also introduce a graph federated learning technique based on learning by contrast. The node classification and graph classification studies show that it outperforms both federated average (FedAvg) and model-contrastive federated learning (MOON)[Zeng, 2022].

Table : Data Privacy in Federated Learning using RAI

A decentralized approach to model training was used in the FL-based method for banking fraud detection, enabling clients to train on their local data and then exchange model updates with a central server. Since raw data stays local while yet gaining from the insights of various datasets, this method puts data privacy first. The project used multithreading to address potential scalability issues and used Flask for server-client communication. To aggregate insights from different models and update the global model, client weightsâ€"multidimensional arrays that represent the neural network's learned parametersâ€"were essential. By incorporating SHAP (SHapley Additive exPlanations) is a popular XAI technique that aims to both provide accurate model predictions and provide insight into the features that have the greatest influence on them. This can be particularly helpful in delicate fields where comprehending the reasoning behind forecasts is just as crucial as the forecasts themselves. Within the framework, SHAP provides a means of fostering trust and guaranteeing that the federated model's decisions are comprehensible and supportive [Awosika, T,2024]. Table shows how RAI ensures privacy of data in Federated Learning

The federated learning community can create and implement AI systems that are ethical, transparent, fair, privacy-preserving, and reliable by incorporating RAI concepts and practices. Realizing the full potential of federated learning in a variety of fields, from financial services to smart transportation, while solving societal issues and fostering public trust, requires this convergence between RAI and federated learning.

## 8. Ongoing research and industry projects

In this section, we have analyzed and evaluated ongoing RAI research and industry projects. Table 6 represents RAI-based research projects and their relevance to numerous technologies and applications.

*8.1. Towards a RAI-driven transformation of work [RESP-AI]*

Recent advancements in technologies such as RAI and Generative AI are generating numerous opportunities and changing workplace dynamics. However, many-AI experts predicted that the latest technologies such as responsive AI can significantly affect the available job roles and employment opportunities. Furthermore, some AI experts suggested that the new AI challenges will also open new horizons for innovation and creativity. AI's future will also depend on the implementation of ethical AI policies within the organization and respective governments. However, the proposed RESP-AI project aims to assist dynamic workplaces in designing a complete roadmap and comprehensive integration of AI technologies within organizations. The proposed project aims to develop skilled individuals using qualitative and quantitative transformative AI methodologies within organizations. The RESP-Ai project has the potential to provide ethical deployment of RAI technology within the organization[92].

*8.2. HumanE-AI-Net Network*

The HumanE-AI-Net project is an ambitious initiative to integrate research organizations, educational institutes, industries, and start-ups in the European Union. The proposed HumanE-AI-Net project aims to create a collaborative network of research scientists, technicians, AI engineers, and computer science experts. Furthermore, the proposed HumanE-AI-Net project is encouraging innovative and creative research projects in domains such as HCI(Human Computer Interface), BCI(Brain Computer Interface), Cognitive and social science, Healthcare, etc. [93]. The proposed HumanE-AI-Net project has discussed various AI-related challenges such as trustworthy and ethical AI systems adaption within real-world industry environments, and complexity issues of RAI-integrated technologies. Furthermore, the proposed HumanE-AI-Net project also provides the foundation for innovative scientific approaches integrated with European ethics and values.

Table 4: RAI Challenges and Solutions

| S. No | Area | Challenge | RAI Solution | Example | Impact of Bias | RAI methods used |
|---|---|---|---|---|---|---|
| [1] | Healthcare | Protecting individual privacy | RAI principles for patient privacy | RAI deployment | Privacy violations, bias lack of transparency | Algorithmic techniques, ethical frameworks |
| [2] | Smart Transportation | Ensuring RAI use | Integrating RAI principles | Federated Learning in Transportation | Potential biases in transportation AI | Analyzing FL roles, promoting effective AI |
| [3] | Data privacy, ML/DL techniques | Ensuring federated learning trustworthiness | Developing trustworthiness taxonomy | FederatedTrust solution | Biases in federated data/model | Taxonomy development, trustworthiness metrics |
| [4] | Finance | Balancing high-performance AI | Leveraging privacy-preserving learning | User-centered federated learning | Privacy and security risks | Federated learning architectures |
| [5] | Graph Data Analysis | Non-IID data distribution | Feature-contrastive graph learning | Graph Federated Learning | Biases in graph data | Contrastive learning methods |
| [6] | Financial Fraud Detection | Balancing transparency | Explainable AI for fraud detection | Federated fraud detection | Biases in fraud detection algorithms | Federated learning, public engagement |

*8.3. Responsible Human-centered AI in Africa [RHAI-AFRICA]*

Recently, AI-related technologies have evolved in developing regions such as Africa. The proposed RHAI-AFRICA project aims to tackle future socioeconomic challenges, integrate responsible-enabled systems within organizations, and design complexities. In low-income developing countries such as Africa, AI technologies can play a vital role in transforming University and school education. AI-enabled educational platforms have the potential to provide theoretical and practical insights about real-world issues. The proposed RHAI-AFRICA project will provide a collaborative opportunity for researchers, academicians, industry personnel, and students to develop an educational ecosystem in Africa. The RHAI-AFRICA project proposed to design a common platform for sharing creative and innovative curriculum design-related content, innovative AI-integrated pedagogical methodologies, and RAI-based educational solutions for students. Furthermore, the proposed project will also emphasize developing sociocultural and ethical policies for AI-integrated educational systems in Africa[94].

*8.4. RAI networks for industries and governments in Latin America*

The proposed RAI Networks project aims to create a collaborative global RAI network for Latin America. The proposed project aims to design standard RAI practices and ethical regulations for the use of RAI systems within government, private, and research organizations. Furthermore, the proposed project will also assist developing countries in designing RAI policies and regulations for the responsible use of AI-based systems. The project also discusses the socioeconomic consequences and opportunities of the use of RAI technologies in various countries that import AI-based technologies and frameworks. Moreover, the proposed project has also discussed the ethical and economical use of RAI frameworks, understanding, and implications of RAI-integrated systems in Latin America [95].

## 9. Challenges and Best practices of RAI

Table : challenges, best practices and benefits

AI technologies have become ubiquitous in various fields like education, manufacturing sector, Agriculture and in health sector. Heuristic search problem, problem solving algorithms, reasoning, self-learning and optimization technique are widely used in all applications directly or indirectly through cloud environment. Most industry and public sectors are applying these techniques for their products and services. Even culture and arts are using AI techniques for implementing their innovative ideas and to develop creative tools. The need for using AI to solve the real world problems has increased and it is beneficial to tremendous applications which impacts socially good. There are many initiatives taken to address the societal needs and to encourage R&D in AI to actively contribute towards the Sustainable Development Goals (SDGs).

Table 5: Research projects on RAI and their relevance to numerous technologies and applications

| Projects (Funding) | Types of Technologies | | | | | Relevant RAI applications | | | | | |
|---|---|---|---|---|---|---|---|---|---|---|---|
| | AI | Machine Learning | IoT | Federated Learning | RAI | Digital Health | Finance | Banking | Agriculture | Education | Transportation |
| RESP-AI (EU) | | ✓ | | | ✓ | ✓ | | | | | |
| HumanE-AI-Net HumanE AI Network | ✓ | | | | | ✓ | ✓ | ✓ | | | ✓ |
| RHAI-AFRICA | | | ✓ | ✓ | ✓ | | | | ✓ | ✓ | |
| RAI networks for industries and governments | ✓ | ✓ | | | ✓ | ✓ | ✓ | ✓ | ✓ | ✓ | ✓ |

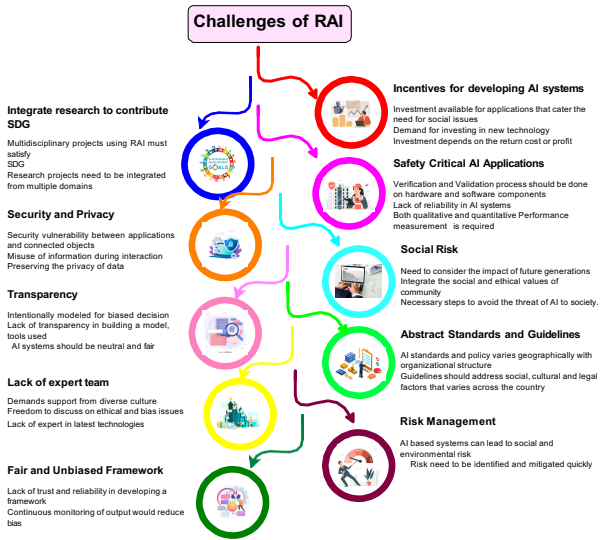

Figure 5: Challenges of RAI

*9.1. Challenges*

**Incentives for developing AI systems:** The main issues that is involved with fostering AI system to be socially fit is providing incentives [96]developing and implementing AI applications. Generally market tends to invest money on high and rapid return on investment. Investors may hesitate to provide research fund to develop socially beneficial applications that uses the technology which is in initial phase. Non-profit organizations and industries have come up to provide fund for the development of AI systems. Also, all the countries have started to have an AI development plan for their requirements.

**Integrative Research to contribute SDGs :** The granularity of academic research in AI tend to focus on specific disciplinary targets. They promote investigations often in vertical domain to foster in-depth and formalized knowledge. This way of carrying out research is certainly needed for the progress of science and research. But certainly this would not be sufficient to amply the AI contributions for socially good. Inorder to develop AI system to substantiate SDG goals, focusing on horizontal and interdisciplinary domain is mandatory. It is not a matter of just applying AI in single domain but is should be a integrative research projects from multiple domain for heterogenous tasks. The main challenge in integrative research is combining data based modelling and model-based reasoning and it highly demands bottom-up learning and rationalization of top-down approach. This heterogeneous task requires multiple functions like sensing, data extraction, reasoning, underlying ontology and to organize and explain the perceived field. Integrative research requires fusing of diverse input sources and integrating multiple knowledge representations and mathematically processing the heterogeneous approaches. To implement this kind of research, it requires participation from all the stakeholders, social actors and contributors. Integrating all the stakeholders is more complicated due to their diverse culture and methodology. But this integration is essential to convert and develop the real issues into relevant contributions which will be assessed mainly by the effective field success.

**Safety critical AI applications:** AI algorithms are frequently applied in various applications which are inbuilt with sensory-motor capabilities and with increasing level of granularity. Components like robots, cyber physical systems, drones and sen-sor components are used in Automated systems integrated with safety critical applications AI system that involve very high environmental and setup costs. For example in health sector, transportation, network management, surveillance and defence systems. As it involves high risk of human life, there should be strict certification procedures that have technical descriptions and some declarations of conformity for correct working of system, complexity and transparency of AI models and in-depth traceability of hardware and software components. The risks in human lives and cost involved in building AI systems are not properly assessed and studied. The main challenge is about the Verification and Validation process on hardware and software components used in AI. It is important to measure and validate the safety of component and systems in AI systems. Any system that uses AI should consider the following points. Formal assumptions about the system and environment, Clear documentation about the functionality and limitation and to determine the basic properties like false positives, reliability and uncertainty of data. Automated hardware and software tools in AI interact with partial or incomplete environment. The main challenges are as follows, measuring the uncertainty of the system that includes nature of data and model used [97], monitoring the functioning of the model with respect to assumptions and environment, assessing the verification and validation properties of complex system with AI algorithms and black box type components. To summarize it is vital to provide awareness to designers who create critical systems about the open challenges and limitations in current techniques[98],[99], mitigating risk and proper vigilance in deployments.

**Security and Privacy:** Currently AI has become an connector between the human and the digital world. There is a rapid growth in interactions between the user to machine with environment is digital through AI. Billions of people are connected through digital world through data, knowledge and based on their semantic content. A vocal assistant that perceives the oral requests from the user in natural language, query engines that interpret each request, processing of video and audio images from various sensors that connected with semantic content are some of the applications where the user interacts with data and model. Even though AI technologies have become a undistinguishable part of digital technologies as per the perceptions of the field, there are uncritical expectations about the correctness of the algorithms and computations, non-reliable decision making, legitimate concerns about the privacy and security of user data and indirect control of human through emotions [100]. Challenges related to security and privacy are listed below

- Security and digital interactions: When AI system is built to interact with the environment with different state transition, there exists an security vulnerabilities between the applications and connected objects. For example in vacuum cleaner robots [101], there exists a high risk of adversarial attacks [102]

- Privacy of data: When there is a transfer of users data for modelling purpose there is a chance that the intruders can misuse the information during interaction of AI system with the environment.

- Intelligibility and transparency: Decision support system should be able to explain the transparently about the assumptions, limitations and boundaries of the system. It should be rational that meets all the requirements of the user and must be able to explain and justify the response to the user request. All the parameters should be easily understandable to the user who uses the AI system.

To summarize, there is a weak regulatory body and economic constraints that pave the way to insufficient deployment with security and trust issues. The contractual relationship [103] between human and a platform is unbalanced that leads to userâ€™s vulnerability to platforms. Services are offered that are regarded as essential to everyone for a modern digital social life, but at a high hidden cost. The UN and EU should provide ethical guidelines [104]but it is currently not sufficient. The immediate requirement is framing more regulations and public policies for avoiding vulnerability and data loss and to maintain security and privacy on userâ€™s information. More awareness and social experiments are needed to support and foster digital regulation on AI automation related to privacy and security of data.

**Social Risk:** The acceptability of technology in public that adopts and uses technology is much more demanding. Social acceptability needs to consider the impact on future generations, should take of social cohesion related to resource sharing and employment. It should also integrate the social and ethical values of community and should consider the global constraints affecting the environment. AI has opened many opportunities in various applications and solved many problems that benefit the human society. But the studies have shown that AI will be a threat to democracy as it will create more risk for human with observed impacts [105] [106]

**Transparency:** Tools used for decision support system may be biased and may not transparent. In come cases the AI engine is intentionally modelled for biased decision to market and recommend a special product for commercial purpose. Customers should be aware of all these marketing strategies and should not become the victims for the business profit. AI systems should be neutral and fair specifically in applications like health, career and legal assessment, insurance, banking and surveillance systems [107]. There is a lack of transparency in training data, building a model, decision making system, environment and the assumptions defined for the model. There need to be more regulations and policies defined for auditing the transparency of the above systems.

To summarize, the need for RAI with respect to social and ethical risk requires various measures to provide standard regulatory mechanisms for benefit of society. It may take quite long time to understand, spread awareness and to build social forces to frame a standard ethical guidelines. But the momentum of AI technology is growing exponentially where it is not able to meet the pace of framing ethical AI guidelines. Even though AI can contribute to development and effectiveness of the society, social experimentation is required before the deployment of AI model as it will reduce the discrepancy between the ethical guidelines and technology momentum.

**Abstract standards and guidelines:** Globally, the guidelines of AI vary culturally, geographically with respect to their organisational structure. These guidelines are not uniformly defined from the authoring bodies and it is often conflicting with the practical use. To design a common guidelines, complete range of stakeholders need to be considered with social experimentation. Even though most of the industry claims that RAI should satisfy fairness, accountable and empower people, practically it has different interpretation while translating these principles for implementation. Most of the definition are directly actionable and has been criticised by many researchers about this issue [108] [109][110]. In general the Principle are state and not defined leaving the room for practitioners discretion to adopt the guideline failing to fulfil the essence of guideline being defined. This diverging interpretation has amplified when it is results in heterogeneous background. Abstract level guidelines need to be converted by practitioners into low level actionable items for the particular use case. They have to consider social, cultural and legal factors that varies across the country, context and between the individuals [111]. Finally with diverse interpretations of RAI principles, the main challenge is choosing the values that are incorporated and the conflicts that exists in different values in specific context

**Lack of Expert Team** Human resources working in AI application generally lack in diverse dimension and critical thinking that leads to blind spots in finding the biases or impacts. While building diverse team to work for AI application, we need to choose not just in terms of location but also from various disciplines. Expertise can be from social expert, ethicists and also good in domain expert. Scaling ethical AI is challenging if the organizations are lacking in appropriate human skills. Industries should focus on providing training on ethics and technology to automate the ethical checks which will facilitate the implementation of RAI effectively. To summarize, the current industry demands support from diverse culture. Also, AI professionals require freedom to discuss on ethical and bias issues. Proper training need to be given for the people to stay updated with the latest technologies.

**Fair and Unbiased Framework** Implementation, guidelines and framework of AI models varies from industry to industry that leads to different perspective of dealing AI responsibly. Opinion on handling AI ethically and legally differs among the organizations which results in unfair and biased framework. So, solution to overcome this challenge is, continuous monitoring of output would reduce the bias and to design a strict ethical guidelines to implement fair AI models to benefit the human society. To summarize, a trustable and reliable framework need to be developed across the firms.

**Risk Management** Organizations and industries are demanding to capitalize and invest on AI technology to avail its benefits. But this rapid adoption is associated with challenges and risks. Some of the risks are aligned with data, model, operations, ethical and legal aspects. Performance of AI application depends on the quality of input data sets used for execution. However data is always vulnerable to data breach, data privacy and threat, data poisoning and cyberattacks. So, data risk can be mitigated by maintaining the data integrity, protecting and securing the data throughout the lifecycle of AI from designing to deployment. Model risk is the treat that involves tampering the weights, architecture and the components of the model. Operational risks are models that are susceptible to model drift, biased nature of framework which leads to failure in system and cybersecurity vulnerabilities. After implementation during developing and deploying AI systems, ethical and legal aspects need to prioritized. If this risk is not mitigated properly then the outcomes produced will be biased and unfair. For example, if the application designed for hiring decision has biased training data, then this might create AI models that is partial to some demographic groups, gender bias or racial type over others. To summarize, type of risk need to be identified throughout the AI lifecycle and need to mitigated quickly.

*9.2. Best Practices*

The governance processes and standards of RAI, should be practiced and repeated systematically while designing the system. The best practices that are followed while implementing Machine Learning project, also holds good for implementing RAI model. Some of the best practices that need to be followed are

- Racially diverse and gender equality should be maintained in the team that works creating RAI standards. Creating a diverse culture would support the team members to speak freely about the bias and ethical issues in AI

- Transparency in creating a explainable AI model need to be promoted. This enables to easily understand the decision made by the AI system

- RAI processes and standards should be measurable as much as possible. When we talk about responsibility it goes with qualitative measure, so there should be some measures to quantify visibility and explainability process in the AI system. May be there can be technical and ethical frameworks to audit the process.

- AI models should be inspected using RAI tools such as TensorFlow toolkit and other tools.

- Metrics need to be identified to monitor and keep track of the errors, biases and false positive cases at minimum.

- Bias testing and predictive testing should be performed that helps to produce the authentic results and trust to the end user.

- Monitoring and Maintenance should be continuous process that goes beyond deployment. This will ensure that the AI system functions in an ethical and unbiased way.

- People should learn from the process and need to stay mindful throughout the process. An organization learns more during implementation of RAI through fairness in following the standards and technical ethics.

Table 6: Challenges and Solutions of RAI

| Sno | Challenges | Solutions |
|---|---|---|
| 1 | Standards and policies varies with organization | Common standards and policies need to be framed to develop a RAI |
| 2 | AI models are not transparent and trustable | Proper documentation using blockchain technology can be implemented to trust the process |
| 3 | AI tools are complex and not known to end-user | Awareness on AI tools can be provided to all the stakeholders |
| 4 | Lack of Reliability in AI models | White box AI model to provide an explainable AI system for each decision made by an AI system. |
| 5 | No proper design and implementation framework for AI | A standard policies for design and implementation process need to be framed for the benefit of human society |
| 6 | Lack of proper expert team to solve the ethical issues | Creating a diverse culture would support the team members to speak freely about the bias and ethical issues in AI |
| 7 | Performance of AI is measured qualitatively | There should be some quantifiable measure to audit the performance of AI models |
| 8 | Challenge in developing a fair and unbiased frameworks | Continously monitoring the output of the model would avoid the AI system working in biased way |
| 9 | It is difficult to understand the decision made by the AI system | AI models developed should be explainable and simple |
| 10 | Risk management in AI systems is difficult | AI risk should be Identified and Mitigated effectively |

## 10. Conclusion

Any organization must adopt AI standards at all levels both in organizational and product levels to ensure risk, security, and trust. The alignment of enterprise with the standards ecosystem would help us to develop an ethical and safe AI system. AI and other recent technologies have tremendous effects and also have less desirable consequences. Social and legal mechanisms of using the technology and the momentum rendered during the deployment of technology make the RAI more challenging. AI scientists and professionals should be more accountable and responsible for providing social awareness about using the AI in an ethical way. Researchers should focus on integrative research that works toward the required paradigm shift to facilitate socially beneficial projects that address the human and social risks of using AI. The development of AI-based projects needs to be aligned with social responsibility and focus on ethical usage of AI that benefits for human community.